\documentclass{aa}  

\usepackage{natbib}
\bibpunct{(}{)}{;}{a}{}{,} % to follow the A&A style

\usepackage{adjustbox}
\usepackage{graphicx}
\usepackage{txfonts}
\usepackage{hyperref}
\hypersetup{
    colorlinks=true,
    citecolor=blue,
    linkcolor=blue,
    filecolor=blue,      
    urlcolor=blue,
    }
\begin{document}

   \title{Galaxy evolution in compact groups. II.  Witnessing the influence of major structures in their evolution}
    \authorrunning {Montaguth et al.}
   \author{Gissel P. Montaguth
          \inst{1,2},
          Antonela Monachesi\inst{1},
          Sergio Torres-Flores\inst{1},
          Facundo A. G\'omez\inst{1},
          Ciria Lima-Dias\inst{1,3},
          Arianna Cortesi\inst{4},
          Claudia Mendes de Oliveira\inst{2},
          Eduardo Telles\inst{5},
          Swayamtrupta Panda\inst{6},
          Marco Grossi\inst{4},
          Paulo A. A. Lopes\inst{4},
          Ana Laura O'Mill \inst{7,8}
          Jose A. Hernandez-Jimenez\inst{9}, 
          D. E. Olave-Rojas\inst{10}
          Ricardo Demarco\inst{11}
          Antonio Kanaan\inst{12},
          Tiago Ribeiro\inst{13},
          William Schoenell\inst{14}
          }

   \institute{Departamento de Astronomía, Universidad de La Serena, Avda. Ra\'ul Bitr\'an 1305, La Serena, Chile
   \and Instituto de Astronomia, Geof\'isica e Ci\^encias Atmosf\'ericas da Universidade de S\~ao Paulo, Cidade Universit\'ria, CEP:05508-900, S\~ao Paulo, SP, Brazil     
         \and Instituto Multidisciplinario de Investigaci\'on y Postgrado, Universidad de La Serena, Avda. Ra\'ul Bitr\'an 1305, La Serena, Chile
        \and Observat\'orio do Valongo, Universidade Federal do Rio de Janeiro, Ladeira Pedro Ant\^{o}nio 43, Rio de Janeiro, RJ, 20080-090, Brazil
        \and Observat\'orio Nacional, Rua General Jos\'e Cristino, 77,  S\~ao Crist\'ov\~ao, 20921-400 Rio de Janeiro, RJ, Brazil
        \and Laborat\'orio Nacional de Astrof\'isica, R. dos Estados Unidos, 154 - Na\c{c}\~oes, Itajub\'a - MG, 37504-364, Brazil
        \and CONICET, Instituto de Astronomía Teorica y Experimental (IATE), Laprida 854, Córdoba X5000BGR, Argentina 
        \and Observatorio Astronómico  de Córdoba  (OAC), Universidad Nacional de Córdoba (UNC), Laprida 854, Córdoba X5000BGR, Argentina
        \and Universidad de Investigación y Desarrollo, Departamento de Ciencias Básicas y Humanas, Grupo FIELDS, Calle 9 No. 23-55, Bucaramanga, Colombia
        \and Departamento de Tecnologías Industriales, Facultad de Ingeniería, Universidad de Talca, Los Niches km 1, Curicó, Chile
        \and Institute of Astrophysics, Facultad de Ciencias Exactas, Universidad Andr\'es Bello, Sede Concepci\'on, Talcahuano, Chile
        \and Departamento de F\'isica, Universidade Federal de Santa Catarina, Florian\'opolis, SC, 88040-900, Brazil
        \and NOAO, 950 North Cherry Ave. Tucson, AZ 85719, United States
        \and GMTO Corporation, N. Halstead Street 465, Suite 250, Pasadena, CA 91107, United States
             }

   \date{Received xxx xx, xxxx; accepted xxx xx, xxxx}

\abstract{Compact groups (CGs) of galaxies are an extreme environment for the morphological transformations and the cessation of star formation in galaxies. However, despite initially being conceived as isolated systems, it is now widely recognised that many of them are not as isolated as expected. Our objective is to understand the dynamics of CGs, as well as how the environment surrounding CGs impacts their morphological and physical properties. To achieve this, we selected a sample of 316 CGs in the Stripe 82 region, with a total of 1011 galaxies, and a sample of 2281 field galaxies as a control sample. We find that at least 41\% of our sample of CGs are part of major structures, i.e. non-isolated CGs. We find a bimodality in the effective radius ($R_e$)-Sérsic index ($n$) plane for all transition galaxies (those with $(u-r) > 2.3$ and $n<2.5$) in CGs. Additionally, transition galaxies in isolated CGs populate more densely the $R_e-n$ plane for $n < 1.75$. In contrast, transition galaxies in non-isolated CGs show a bimodal distribution in the $R_e-n$ plane, with the $n$ values smoothly increasing towards higher values, and 62\% of these galaxies having $n > 1.5$. This indicates that the majority of these galaxies have already undergone a morphological transformation and primarily contribute to the population of more compact galaxies in the $R_e-n$ plane.
We find that galaxies in our sample of CGs have a lower mean specific star formation rate (sSFR) compared to the control sample, with non-isolated CGs showing even lower sSFR values, indicating that dense environments suppress star formation. Additionally, non-isolated CGs have a higher fraction of quenched galaxies relative to isolated CGs and the control sample. When dividing the sample by morphology, we find significant differences only for early-types galaxies (ETGs; those with $(u-r) > 2.3$ and $n>2.5$). In isolated CGs, ETGs show a lower fraction of quenched galaxies and higher sSFR in low-mass bins ($\log(M_*/M_{\odot}) < 11$), suggesting they can maintain star formation. In contrast, ETGs in non-isolated CGs exhibit a higher fraction of quenched galaxies and lower sSFR in high-mass bins ($\log(M_*/M_{\odot}) > 11$), underscoring the environmental impact on star formation suppression. Based on our results, we propose an evolutionary scenario where the major structures in which the CGs are embedded accelerate the morphological transformations of their galaxy members, and also facilitates preprocessing.
Our findings highlight the importance of considering the larger structures in which CGs may be located, when analysing the properties of their galaxy members, as this can significantly affect the evolution of CGs and their galaxies.
}

   \keywords{galaxies: groups: general - galaxies: evolution - galaxies: interactions}

   \maketitle
%
%-------------------------------------------------------------------

\section{Introduction}

Galaxy groups and clusters play a significant role in helping us understand how the surrounding environment shapes the evolution of galaxies. In the $\Lambda$-CDM model (\citealt{1984Peebles}), galaxy clusters mainly grow via the accretion of galaxies, both individually and in groups that have already undergone environmental effects, known as pre-processing (\citealt{1998Zabludoff}, \citealt{2004Fujita}, \citealt{2019MNRAS.488..847P}, \citealt{2021Donnari}), which has been supported by observations of groups falling into clusters (\citealt{2014A&A...570A.119E}, \citealt{2015Haines}, \citealt{2024Lopes}). Compact groups (CGs) of galaxies, characterised by high densities comparable to cluster cores and a low-velocity dispersion, provide an excellent opportunity to study the effects of galaxy-galaxy interactions (\citealt{hickson1982systematic}). Observational studies and simulations suggest that CGs consist of a combination of virialised groups (\citealt{1997Gomezfl}, \citealt{2001GomezFle}), chance alignments, where they are merely a projection effect (\citealt{1986Mamon}, \citealt{1995Hernquist}, \citealt{2006Tovmassian}, \citealt{2020Hartsuiker}), collapsed groups, which refer to bound substructures within clusters, and nodes (\citealt{2010Diaz}, \citealt{2021Zheng}, \citealt{2022Taverna}, \citealt{2023Taverna}). This shows that the physical nature of CGs is still not well understood, which is a challenge at the observational level as there is no access to the 3D information of each CG.

CGs have been classified in three main types by \cite{2004Coziol}, based on their internal dynamic and galaxy properties. Type A CGs are predominantly characterised by low-velocity dispersions with a median of 135 $km/s$, and contain a large fraction of late-type spirals with active star formation or active galactic nucleus. Type B CGs have intermediate velocity dispersions with a median of 302 $km/s$, and consist of a significant fraction of interacting or merging galaxies. Type C CGs have high velocity dispersions with a median of 567 $km/s$, and are dominated by elliptical galaxies that are passive. These authors suggest that CGs evolve from Type A to Type C, proposing that the evolutionary state of a CG increases with its velocity dispersion. Additionally, the environment surrounding the sample of CGs studied by \cite{2004Coziol} was analysed by other authors.

For example, \cite{1998Ribeiro} found that these CGs could be either locally isolated or surrounded by galaxies, suggesting that these CGs are part of major structures. Therefore, \cite{2004Coziol} suggests that the evolution of CGs not part of major structures must be regulated by their own masses, while those linked to major structures depend on the masses of the major structures, with the most dynamically evolved CGs found in the most massive structures. \cite{2004Coziol} propose that this may result from galaxies in CGs evolving more rapidly when these groups are located within massive structures. Alternatively, it is suggested that the formation of CGs embedded in massive structures may have preceded the formation of isolated CGs.

The more dynamically evolved CGs have higher velocity dispersion and a lower fraction of late-type galaxies (LTGs) than less evolved systems (\citealt{hickson1982systematic}, \citealt{2020Moura}, \citealt{diaz2021compact}). Within this context, it is worth highlighting a recent study done by \cite{2021Zheng}, where they found that the CGs in their sample are not necessarily isolated; some of them are embedded in major structures, others called `predominant CGs', meaning, groups that have more galaxies in addition to those in the CG catalogue, but these additional galaxies are fainter than the CG members. Others are called `split CGs' because part of their members belong to at least two different groups. One of the most significant findings in that work is that the dynamics of CGs within major structures are influenced by the mass of these major structures. In this context, understanding the properties and dynamics of CGs and their connections to larger-scale structures is crucial for interpreting the environmental effects on galaxy evolution in these systems, as suggested by \cite{2015Diaz} and \cite{2023Taverna}.

In \cite{Montaguth2023} we published the first of a series of papers, aimed at understanding better the role of CGs in the pre-processing of galaxies. In that first work (\citealt{Montaguth2023}) we estimated the Sérsic index and the effective radius of each galaxy, in a sample of 340 CGs and a control sample of field galaxies by modelling the galaxies' surface brightness. We found that galaxies in CGs have smaller effective radii compared to the galaxies in the control sample, suggesting possible influences from tidal interactions. Additionally, we classified galaxies based on their morphology using the colour $(u-r)$ and Sérsic index ($n$). Specifically, we defined `transition galaxies' as those with a lower Sérsic index ($n<2.5$) and red colours ($(u-r)>2.3$). Upon comparing these galaxies between CGs and the control sample, we identified a peculiar population of galaxies that was absent in the control sample. These CG galaxies are characterised by their smaller radius and larger $n$ compared to their counterparts in less dense environments, indicating an ongoing morphological transformation. This finding agrees with previous works, such as those done by \cite{2016Bitsakis} and \cite{2012Coenda}, which have also noted the effective transformation of spiral galaxies into early-type galaxies within CGs.

Moreover, we found that CGs hosted a higher fraction of quenched galaxies than the control sample. Additionally, variations in star formation rates were evident, suggesting that environmental effects favour the cessation of star formation, in agreement with previous studies (\citealt{2007AJohnson}, \citealt{2008Gallagher}, \citealt{2010Walker}, \citeyear{2013walker}). In the first paper, we studied all the CGs together. However, one clue to understanding the evolution of galaxies in these CGs lies in studying their surrounding environment. Considering this information, the following questions naturally arise: are the CGs responsible for the morphological transformation of galaxies that we find in \cite{Montaguth2023}? Or is the environment in which the CGs are embedded which generates this transformation?  How significant is the impact of the major structure where the CGs are located on the physical transformation of galaxies?

Our goal here, in the second paper of this series, is to understand how the dynamical state of a CG is affecting the evolutionary stage of its members and to discern what is the role of the larger-scale environment in the evolution of CG of galaxies. To achieve this, we will use the morphological parameters calculated in \cite{Montaguth2023} and complement them with data from the GALEX-SDSS-WISE LEGACY CATALOGUE (GSWLC, \citealt{2018GSWLC}) which provides star formation rate information. The outline of this paper is as follows: we describe the data used, our process for selecting CGs and control field samples in Section \ref{sec:data}; our methodology to classify galaxies according to their morphology, the dimensionless crossing time, and the compactness of each CG is described in Section \ref{sec:methodology}. We present and discuss our results in Sections \ref{sec:results} and \ref{sec:discussion}; and finally, our conclusions and summary are laid out in Section \ref{sec:conclusions}. Throughout this paper we have adopted a flat cosmology with $H_0 = 70 km$ $s^{-1}$ $Mpc^{-1}$, $\Omega_M = 0.3$, and $\Omega_\lambda = 0.7$ (\citealt{2003Spergel}).

\section{Data}
\label{sec:data}

In this Section, we briefly describe our criteria for selecting the CGs and their galaxies, and the control sample of field galaxies. To identify CGs, we used \cite{zheng2020compact} catalogue. In this catalogue, \cite{zheng2020compact} used the redshifts from Sloan Digital Sky Survey (SDSS) data release 14 (DR14) (\citealt{2018Sdss14}), Large sky Area Multi-Object fiber Spectroscopic Telescope (LAMOST) (\citealt{2015Lamost}), and Galaxy And Mass Assembly (GAMA) (\citealt{2015GAMA}). For the CGs selection criteria, the \cite{zheng2020compact} catalogue uses a combination of Hickson's photometric criterion and a spectroscopic approach that includes velocity constraints between the galaxies ($v_i$) and their group ($v_G$), $|v_i-v_G|\leq$1000 km s$^{-1}$. The photometric criterion considers the group's surface brightness, which is determined by the total magnitude of all galaxies averaged over the smallest circle that encloses them. This value must be lower than 26 mag arcsec$^{-2}$ in the $r$-band. Additionally, the isolation criterion requires $\theta_N \geq 3\theta_G$, where $\theta_G$ is the radius of the group and $\theta_N$ is the radius of the brightest galaxy closest to the group. Unlike the original criteria of \cite{hickson1982systematic}, this version does not include the magnitude difference criterion of $\Delta \text{mag}_r \leq 3$ from the brightest galaxy.

From this catalogue, we identified CGs in the Stripe 82 region (\citealt{2009Dr7}) and obtained the coordinates, redshift of each galaxy, and velocity dispersion of each CG. As we mentioned in \cite{Montaguth2023}, we selected galaxies in this region because it is associated with the first data release of the Southern Photometric Local Universe Survey (S-PLUS) project (\citealt{mendes2019southern}), an ongoing imaging survey that uses a robotic 0.8m aperture telescope located at the Cerro Tololo Inter-American Observatory (CTIO) in Chile. The S-PLUS survey employs the Javalambre 12-band magnitude system, consisting of 5 broad-band filters ($u$, $g$, $r$, $i$, $z$) and 7 narrow-band filters centred on notable stellar spectral features, and it has been used to study the morphological and physical properties of galaxies in clusters (\citealt{2021ciria},\citeyear{2024ciria}) and in CGs in our first work of \cite{Montaguth2023}. We note, however, that the analysis shown in this paper is based on the DR3 of S-PLUS (\citealt{almeida2021data}), where the photometry has been improved.

We then cross-matched the coordinates from \citeauthor{zheng2020compact} catalogues with the S-PLUS catalogue, and we obtained a total of 316 CGs with 1011 galaxies, within a redshift range from $0.015$ to $0.197$. For further details about the S-PLUS survey and its photometry, we refer the reader to \cite{mendes2019southern} and \cite{almeida2021data}, respectively.

For the control sample of field galaxies, we used the catalogue by \cite{2007Yang}, which is based on SDSS-DR7\footnote{The catalogues by Yang et al. are available at: \url{https://gax.sjtu.edu.cn/data/Group.html}}. In this catalogue, the structures are classified based on the halo-based group method, which employs an iterative approach to estimate the mass of the group. In short, it starts by estimating the luminosity and centre of each group using the Friends-Of-Friends (FOF) algorithm (\citealt{1985Davis}). Then, it calculates the average mass-to-light ratios of the groups, initially assuming $M/L_{19.5}=500hM_{\sun}/L_{\sun}$ for the first iteration, where $L_{19.5}$ represents the luminosity of all group members with $M_r-5log(h) \leq -19.5$ and $h=H_0/(100 km s^{-1} Mpc^{-1})$. Subsequent iterations use ratios from the previous step. This mass estimation is used to derive the size and velocity dispersion of the hosting halo, with the halo size determined by the radius $r_{180}$ where the average mass density is 180 times that of the Universe's average density at a given redshift. These parameters are then used to determine group membership in redshift space, and if a new member is identified, the process iterates to reevaluate the group's centre and other parameters.

\begin{figure}
    \centering

    \includegraphics[scale=0.4]{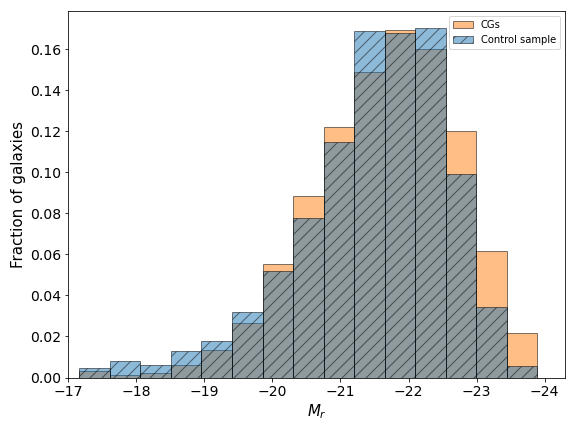}

    \caption{Normalised distribution of rest-frame absolute magnitude in $r$-band for galaxies in CGs (orange), and control sample(blue). The absolute magnitudes in the r-band are corrected for Galactic extinction and include the k-correction.}
    \label{fig:Mr}
\end{figure}

From this catalogue we selected the galaxies that were labelled as single-member groups, meaning, one single galaxy\footnote{In fact, we selected the subsample from that field galaxy sample in Table 3, and we are simply maintaining the nomenclature given by \cite{2007Yang} to the field galaxies}, and that are in Stripe 82. We selected the coordinates, and then crossmatched again with the S-PLUS catalogue, finding a total $11841$ galaxies. From this sample, we used a Monte Carlo algorithm to obtain a subsample of 2281 galaxies, ensuring they share the same range in apparent magnitude in the $r$-band and redshift range as the galaxies in CGs. In Figure \ref{fig:Mr}, we show the distribution of absolute magnitudes in the $r$-band for both samples, with galaxies in CGs shown in orange and the control sample in blue. Further details on the sample selection process can be found in \cite{Montaguth2023}.

\section{Methodology}
\label{sec:methodology}

\subsection{Dynamical state of the CGs: dimensionless crossing time and compactness}

One convenient method for estimating the dynamical state of a group is by using the dimensionless crossing time, $H_ot_c$. It corresponds to the period of time it takes for a galaxy to pass through a group (\citealt{2020Moura}), where $H_o$ is the Hubble-Lemaitre constant. The crossing time, $t_c$, as defined by \cite{1992hickson} is estimated as follows:

\begin{equation}
    t_c=\frac{4R}{\pi V}
\end{equation}

where $V$, the three-dimensional velocity dispersion, also referred to as the 3D velocity standard deviation, is defined as $V=[3(\langle v^2\rangle -\langle v\rangle^2-\langle \delta v^2\rangle)]^{1/2}$, where $\langle v\rangle$ is the average of the radial velocities of galaxies in the group, and $\langle \delta v^2\rangle$ is the average of velocity errors squared, which was estimated using the estimated errors in the redshift. We use the value for radial velocities and their errors already obtained in the catalogues of the CGs. $R$ represents the median of the two-dimensional galaxy-galaxy separation vector in kpc. We estimated it by first calculating the separation in angular distances (in arcsec) between each galaxy in the CGs. Then, we determined the median of these separations in arcseconds for each group and converted it to linear distances (in kpc), according to the distance of each group.

We also estimate the compactness parameter, which was defined by \cite{hickson1982systematic} as the surface brightness of the group. This is the total magnitude of the group galaxies averaged over the smallest circle containing the galaxies:

\begin{equation*}
    \mu = -2.5log  \left( \frac{\sum_{i=1}^N 10^{-0.4m_i}}{\pi \theta_G^2}  \right)
\end{equation*}

where $m_i$ is the apparent magnitude in $r$-band of each galaxy in the group, and $\theta_G$ is the angular radius of the smallest circle, in arcseconds, which contains all the galaxies within the group. We consider this parameter with the purpose of identifying whether the compactness of CGs has any relation to their dynamical state.

\cite{zheng2020compact} used the gapper estimator proposed by \cite{1990AJ....100...32B}, who showed that the standard deviation is not a robust measurement for the velocity dispersion of galaxy groups with small numbers of members. Therefore, we kept the $\sigma_G$ values for the 316 CGs from the catalogue of \cite{zheng2020compact}. According to eq.(12) in \cite{1990AJ....100...32B}, the rest-frame velocity dispersion was estimated as:

\begin{equation}
    \sigma_{G} = \frac{\sqrt{\pi}}{(1+z_g)N(N-1)} \sum_{i=1}^{N-1} w_i g_i,
\end{equation}

where $z_g$ is the CG redshift and $w_i$ is the Gaussian weight defined as $w_i = i(N-i)$. Here, $N$ represents the number of members in each CG. The gaps $g_i$ are defined as:

\begin{equation}
    g_i = V_{i+1} - V_i,
\end{equation}

where $V_i$ is the radial velocity of the $i$-th galaxy within the CG.

\subsection{Morphological classification} 
\label{subsec:Morpho}

The key parameters used to analyse galaxy morphology, the Sérsic index ($n$) and the effective radius ($R_e$), are obtained for each galaxy in our sample and for each available filter using the \textsc{MegaMorph} code (\citealt{2011Bamford,2013Haussler,2013vika}). This code extends the \textsc{GALFITM} algorithm, which models a galaxy's surface brightness using a two-dimensional analytical function, to multiple wavelengths via Chebyshev polynomials. The best fit is determined using the Levenberg-Marquardt algorithm, which minimises $\chi^2$. We fit all galaxies with a single-component model based on a Sérsic profile (\citealt{Sersic}), where the Sérsic index characterises the shape of the light profile.

In this study, we aim to explore the connection between galaxy dynamics and physical/structural parameters in CGs. Therefore, we focus only on the $R_e$ and $n$ values in the $r$-band, since \cite{Montaguth2023} already examined the relationship between structural parameters and wavelength. However, incorporating multiple wavelengths in the fitting process improves the precision of these parameters. As demonstrated by \cite{2013vika}, using \textsc{MegaMorph} with SDSS filters ($u$, $g$, $r$, $i$, $z$) significantly reduces both statistical and systematic uncertainties in structural parameter measurements compared to \textsc{GALFIT}, which performs the fit filter by filter (\citealt{2002Peng}, \citealt{2010Peng}).

Considering that we are using S-PLUS images, which span a wide range of S/N from 10 to 1600 in the $r$-band, this may affect the derivation of structural parameters. Nonetheless, the authors of \textsc{GALFIT}\footnote{Subsection: How biased are the size and luminosity measurements by low signal-to-noise when you can't see outer regions of galaxies? \url{https://users.obs.carnegiescience.edu/peng/work/galfit/TFAQ.html}} emphasise that uncertainties in the fittings should be considered when interpreting structural parameters. They simulated galaxies with different Sérsic profiles and added noise to see if GALFIT could recover the original values used to model the galaxies. They found that in the case of galaxies with higher noise, the original value can be recovered if the uncertainties from the fit measurements are considered.

\begin{figure}
    \centering
    \includegraphics[scale=0.4]{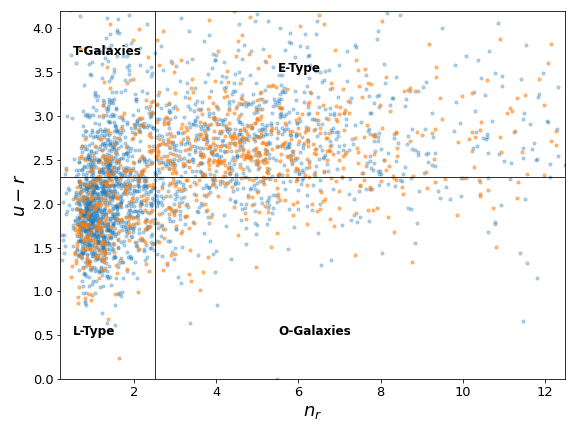}
    \caption{Morphology classification using $(u-r)_0$ rest-frame colour and Sérsic index in the $r$-band. The vertical line indicates $n_r = 2.5$, and the horizontal line indicates $(u-r)_0 = 2.3$. This classification was proposed by \cite{vika2015megamorph}. The blue dots represent galaxies in CGs, while the orange dots represent galaxies in the control sample.}
    \label{fig:vika}
\end{figure}

In this work, we classify galaxies according to their morphology, as defined in \cite{Montaguth2023}. This classification is based on the Sérsic index in the $r$-band ($n$) and the $(u-r)$ colours, following the criteria proposed by \cite{vika2015megamorph}. In contrast to \cite{Montaguth2023}, here we have applied the k-correction to the colour $(u-r)$. Therefore, the colour $(u-r)_0$ used for the morphological classification is k-corrected. We made this correction because \cite{vika2015megamorph} show that this technique remains effective up to z$\sim0.1$, while our sample extends up to  z$\sim0.2$. We employ the publicly available software package of \cite{Blanton2007} in version $V4\_3$ to obtain de-reddened model magnitudes at $z = 0$. Figure \ref{fig:vika} shows the k-corrected colour $(u-r)_0$ - $n$ diagram. The horizontal line represents the colour $(u-r)_0=2.3$ and the vertical line represents a Sérsic index $n=2.5$. The lower-left quadrant defines our selection of late-type galaxies (LTG), the upper-left quadrant, contains what we name transition galaxies, the upper-right quadrant, early-type galaxies (ETGs), and galaxies in the lower-right quadrant are named as `other galaxies'.

As mentioned in Section \ref{sec:data}, the \cite{zheng2020compact} catalogue does not follow the magnitude restriction proposed by \cite{hickson1982systematic}. However, this catalogue specifies which CGs meet this magnitude restriction. Therefore, these differences between the CGs in the catalogue may introduce a potential bias in the morphological composition of the CGs.

\begin{figure}
    \centering
    \includegraphics[scale=0.4]{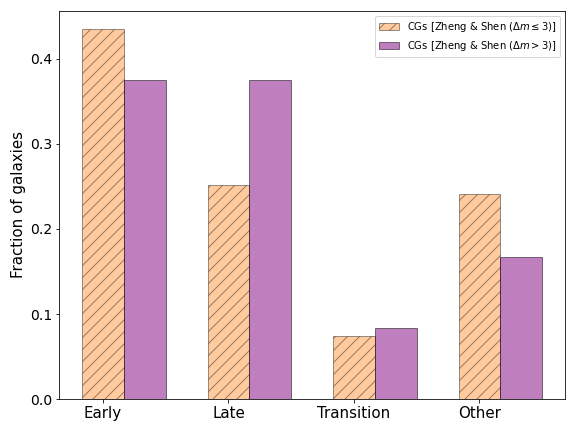}
    \caption{Fraction of galaxies in different morphological categories: ETGs, LTGs, Transition, and Other galaxies. The normalisation of the histograms was performed using the total number of galaxies in their respective samples. The orange histogram, with a diagonal hatch pattern (\texttt{//}), corresponds to CGs from the \cite{zheng2020compact} catalogue with a magnitude gap $\Delta m \leq 3$. Finally, the purple histogram, represents CGs from the \cite{zheng2020compact} catalogue with a magnitude gap $\Delta m > 3$.}
    \label{fig:frac_catalogos}
\end{figure}

However, of the 316 CGs in our study, only 14 of the 316 CGs from \cite{zheng2020compact} do not meet the magnitude restriction, representing just 4\% of the total sample. As shown in Figure \ref{fig:frac_catalogos}, in the case of the CGs that do not follow the magnitude criterion (see the purple histogram), there is a similar fraction of LTGs and ETGs at approximately $0.37$. In contrast, CGs that meet the criterion contain a higher fraction of ETGs at $0.43$ (see the orange histogram with a diagonal hatch). These values were normalised by the total number of galaxies within this subsample. 

The small proportion of CGs that do not follow the magnitude restriction from \cite{zheng2020compact} suggests that our findings are statistically robust across the entire sample. Thus, while some biases may exist, they do not affect the general trends observed in our study.

\subsection{Star formation rate and stellar masses} 
\label{subsec:ssFr_mass}

We complement the morphological classification with information on the star formation rate (SFR) and stellar mass for each galaxy. In the case of the SFRs, we used the values published by \cite{2018GSWLC} who derived the SFRs through spectral energy distribution fitting on GALEX, SDSS, and WISE data. These authors use the \textsc{CIGALE} code (\citealt{2009Noll}) to perform a SED fitting to each galaxy. In this catalogue, we find information for $99\%$ of the sample in CGs, corresponding to 911 galaxies, and $92\%$, equivalent to 2099 galaxies, in the control sample. To estimate the stellar mass ($M_*$), we use an empirical relation proposed by \cite{2011Taylor}. This empirical relation is based on the rest-frame colour $(g-i)_0$ and the absolute magnitude in the $i$-band for each galaxy (for a more detailed estimation see \citealt{Montaguth2023}). We note that this stellar mass estimate is consistent with those obtained from SED fitting by \cite{2018GSWLC}, within a difference in dex of  $0.13$. We decided to use the colour estimated stellar mass because in \cite{Montaguth2023}, we analysed the mass-size relation of all samples of CGs and the control sample. Using these parameters we derived the specific star formation rate (sSFR) for the galaxies in our samples that have the SFR values in \cite{2018GSWLC} catalogue.

\section{Results}
\label{sec:results}

\subsection{Dynamical status of the CGs}
\label{sec:dynamical}

\begin{figure}
    \centering
    \includegraphics[scale=0.4]{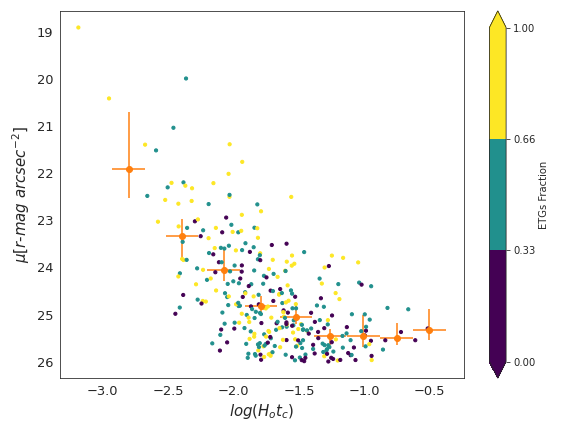}

    \caption{Dimensionless crossing time versus compactness for CGs. The points represent the values for each CG, with the colour of the points indicating the fraction of ETGs. The orange circles with bars represent the median values per bin of crossing times, with the bars on the x-axis indicating bin sizes and those on the y-axis showing the $90\%$ confidence interval.}
    \label{fig:compactness}
\end{figure}

In Figure \ref{fig:compactness} shows the compactness as a function of the dimensionless crossing time. The points represent the values for each CG, with the colour of the points indicating the fraction of ETGs, and the orange circles with error bars represent the median values, where the bar on the x-axis is the size of the bin used to estimate the median and the length of the y-axis is the statistical error of a $90\%$ confidence interval (CI) using bootstrapping. We find that CGs with fainter compactness exhibit higher dimensionless crossing times, which suggests that they are dynamically less evolved than bright CGs.

\begin{figure}
    \centering
    \includegraphics[scale=0.4]{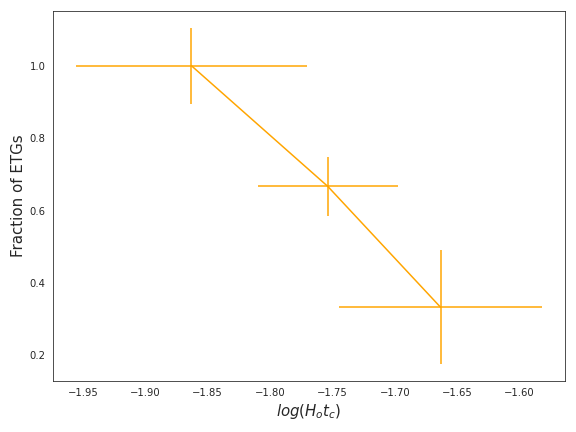}
    \includegraphics[scale=0.4]{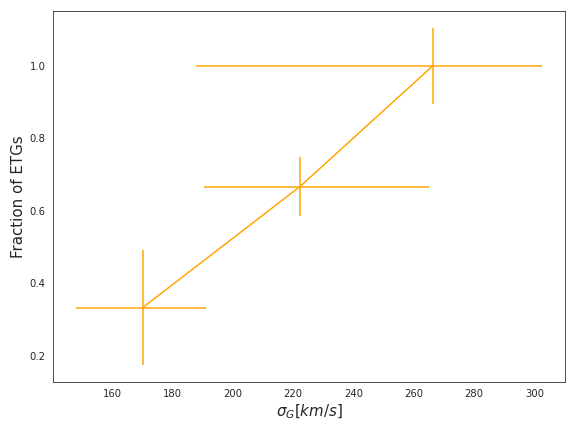}
    \caption{Median fraction of ETGs as a function of the median crossing time (top panel) and velocity dispersion (bottom panel). The median values are estimated from three bins for the ETG fraction: 0-0.33, 0.33-0.66, and 0.66-1. For each $x$-bin, we identified all CGs and estimated the median fraction of ETGs, along with the median values of $\sigma_G$ and $H_{o}t_{c}$. The error bars indicate the statistical uncertainty of a $90\%$ confidence interval, which was determined using bootstrapping.}
    \label{fig:sigma}
\end{figure}

Figure \ref{fig:sigma} presents the median fraction of ETGs as a function of the median crossing time (top panel) and velocity dispersion (bottom panel), respectively. To estimate these median values, we considered three bins for the ETG fraction: 0-0.33, 0.33-0.66, and 0.66-1, as the majority of our CGs have between 3 and 4 members. We selected all CGs within each bin and estimated the median fraction of ETGs, as well as the median $\sigma_G$ and $H_{o}t_{c}$. The error bars represent the statistical error of a $90\%$ confidence interval using bootstrapping.

According to the top panel of Figure \ref{fig:sigma}, the median fraction of ETGs shows an inverse correlation with $H_{o}t_{c}$, indicating that CGs with a higher fraction of ETGs are more dynamically evolved due to their lower $H_{o}t_{c}$ values. In the bottom panel, we observe that the fraction of ETGs increases with $\sigma_G$, which is expected given the relationship between $H_{o}t_{c}$ and $\sigma_G$. This is in agreement with the results presented by \cite{2020Moura}, who divided their galaxy sample into two morphological types: elliptical or spiral. Hence, a high fraction of spirals would indicate a low fraction of ellipticals. They found that when dividing the CGs into low-$\sigma_{G}$ (high-$\sigma_{G}$) with $\sigma_{G} \leq 180$ km/s ($\sigma_{G} > 180$ km/s), the CGs with low-$\sigma_{G}$ have larger crossing times and higher fractions of spirals compared to those with high-$\sigma_{G}$.

In summary, we find that parameters such as crossing time, velocity dispersion, compactness, and ETG fractions provide clues about the evolutionary stage of a CG. Indeed, CGs having low-velocity dispersion, faint compactness, and a lower ETG fraction display high crossing time values, suggesting that they are at an early stage of interaction, or just formed (\citealt{2020Moura}). In general terms, these systems are less evolved than those CGs with higher velocity dispersion, brighter compactness, and higher fraction of ETGs.

\subsection{Connection between velocity dispersion and morphological transformation}
\label{sec:Disperson_mor}

In \cite{Montaguth2023}, we discovered a bimodality in the distribution in the $R_e-n$ plane for transition galaxies in CGs. We found a population of peculiar galaxies that are more concentrated ($n>1.7$) and smaller ($R_e<6 kpc$) compared to transition galaxies in the control sample. Previous studies have shown a strong relationship between galaxy morphology and velocity dispersion in CGs \citep{1988Hickson, 2004Coziol}, with high-velocity dispersion CGs containing fewer LTGs. This relationship is more significant in CGs than in cluster or loose groups \citep{1986Mamon}, and velocity dispersion is a reliable proxy for the group's mass.

\begin{figure*}
    \centering
    \includegraphics[scale=0.5]{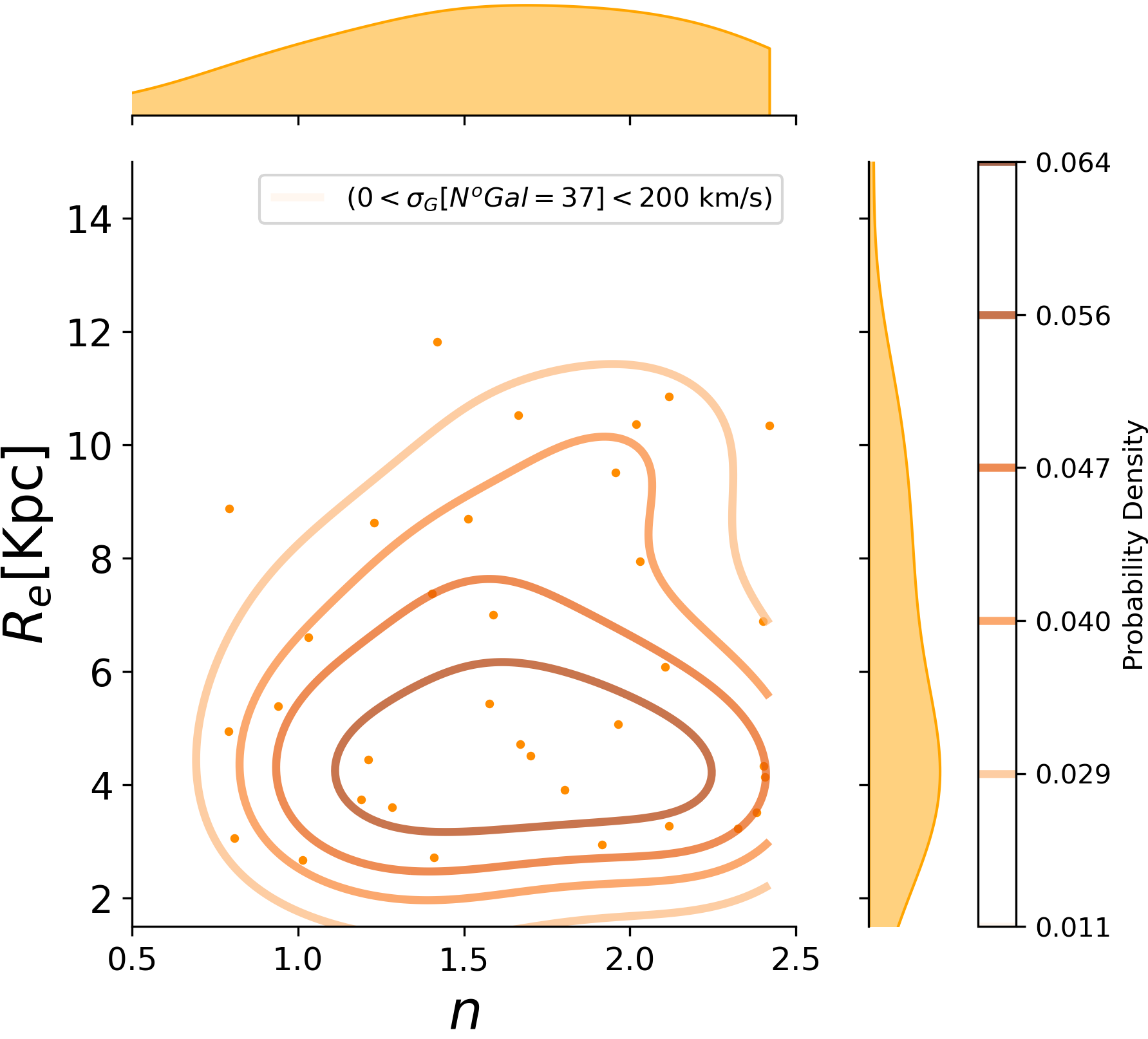}
    \includegraphics[scale=0.5]{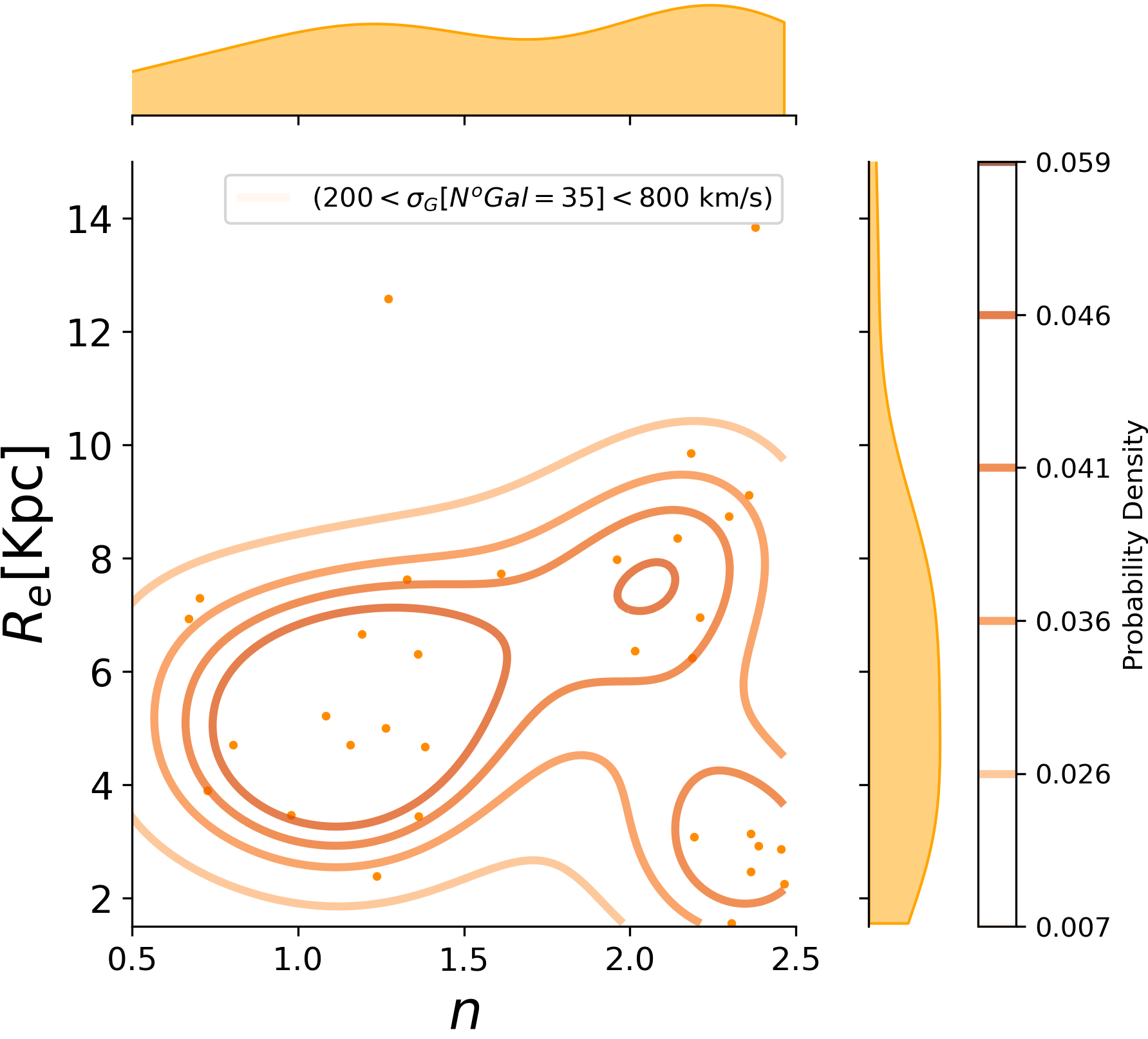}
\\
    \includegraphics[scale=0.5]{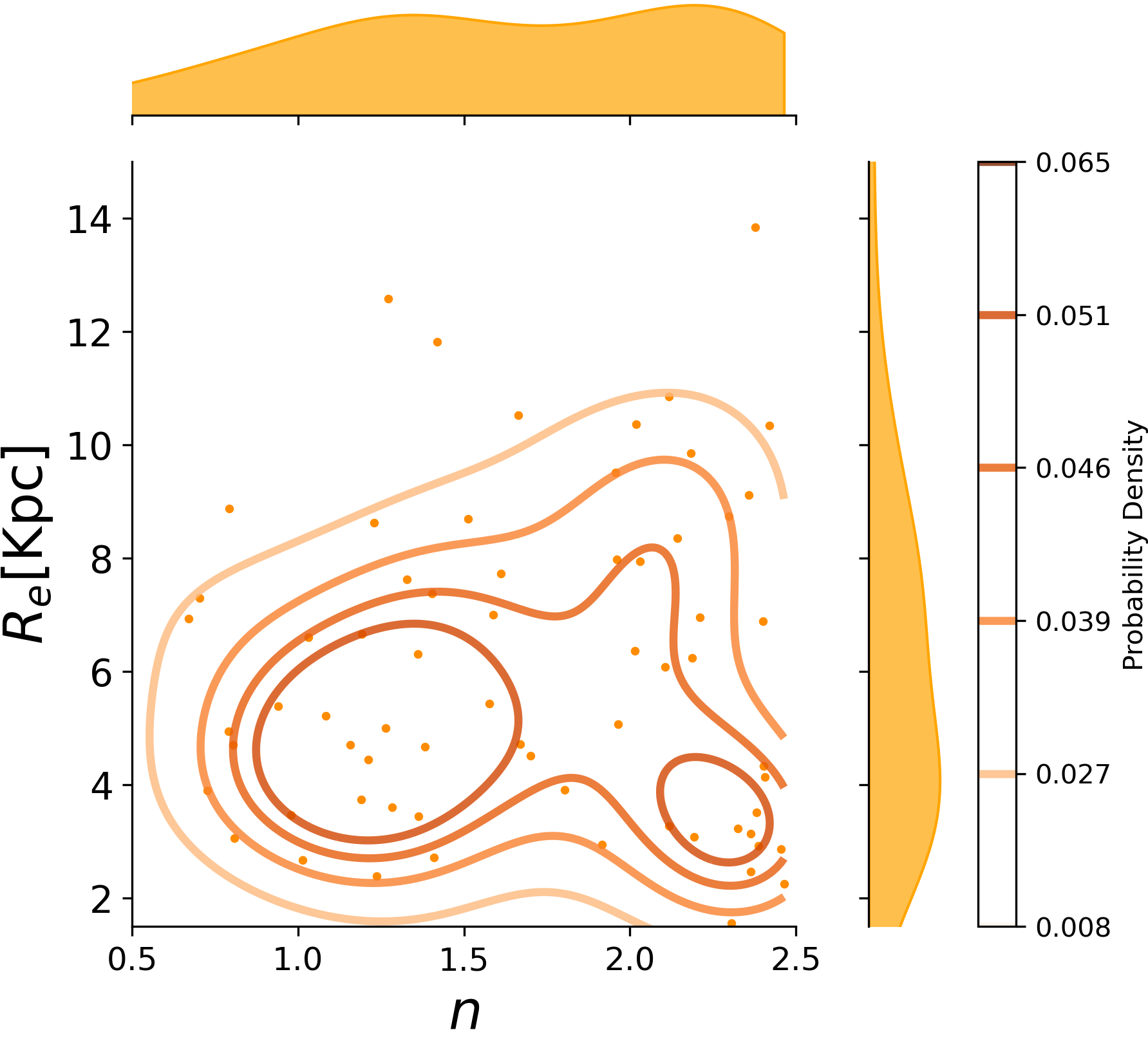}
    \includegraphics[scale=0.5]{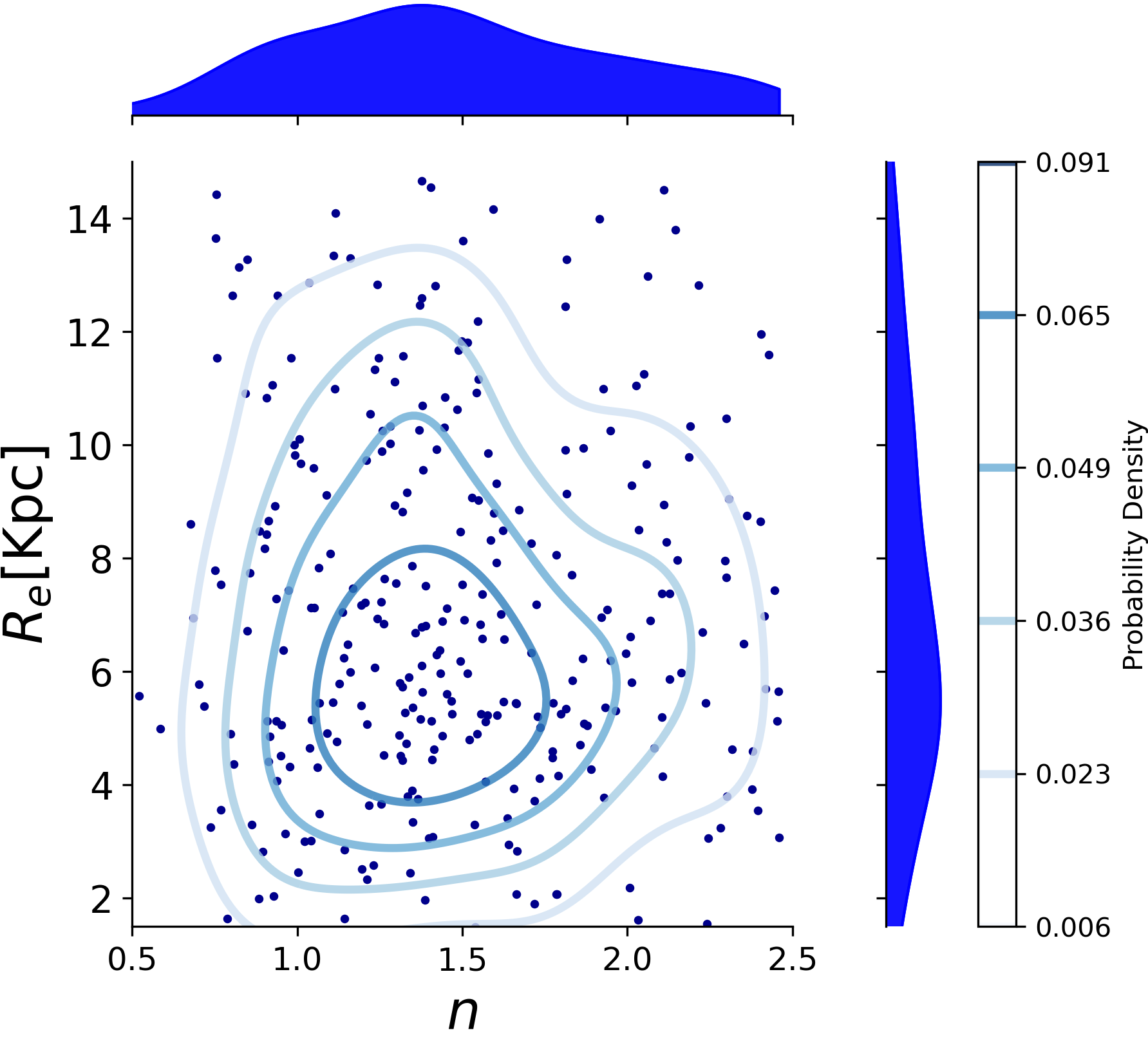}
    \caption{Effective radius–Sérsic index plane for transition galaxies in the $r$-band. Top panels show orange contours for galaxies in CGs, divided by velocity dispersion bins, with the number of galaxies indicated above each panel. Bottom panels display transition galaxies for all CGs (orange contours, left) and for the control sample (blue contours, right).}
    \label{fig:Re_n_sigma_tran}
\end{figure*}

Figure \ref{fig:Re_n_sigma_tran} illustrates the $R_e$ as a function of $n$ for 67 transition galaxies in CGs. The top left panel displays galaxies in CGs with velocity dispersions between $0-200$ km/s, while the top right shows galaxies in CGs with velocity dispersion between $200-800$ km/s. The bottom left panel includes transition galaxies for all CGs, and the bottom right provides a comparison with the control sample.

\begin{figure*}
    \centering
    \includegraphics[scale=0.5]{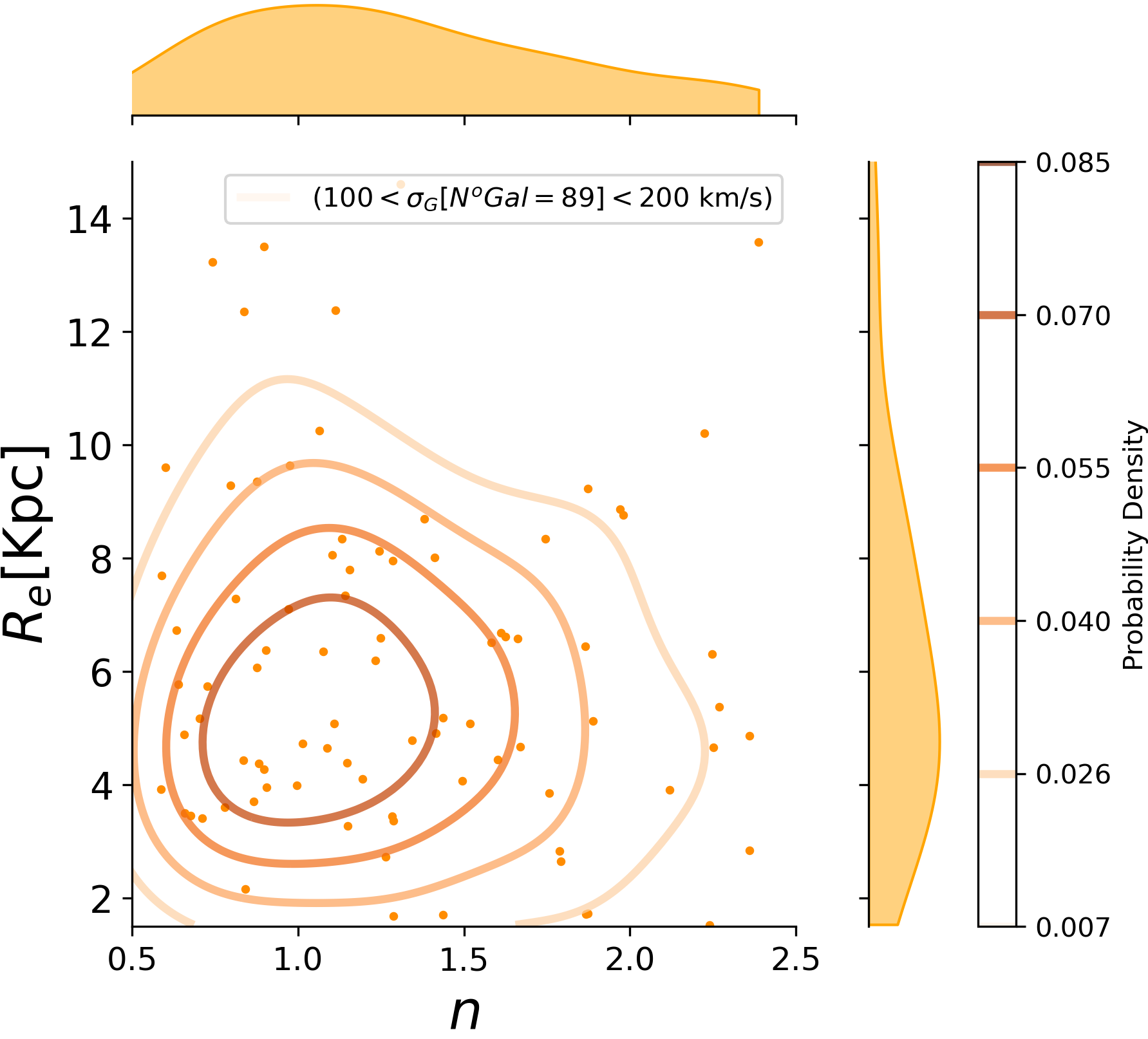}
    \includegraphics[scale=0.5]{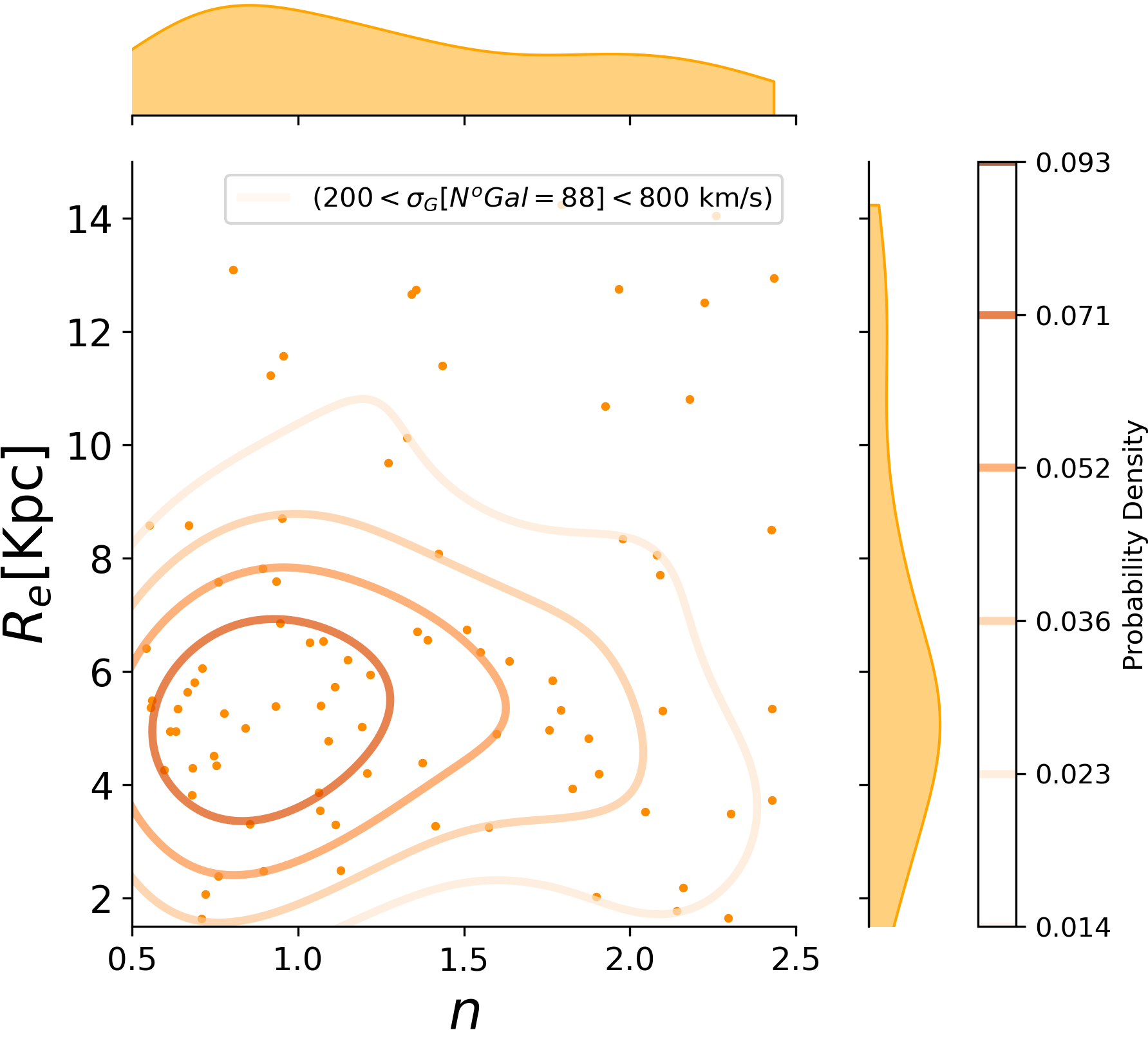}\\
    \includegraphics[scale=0.5]{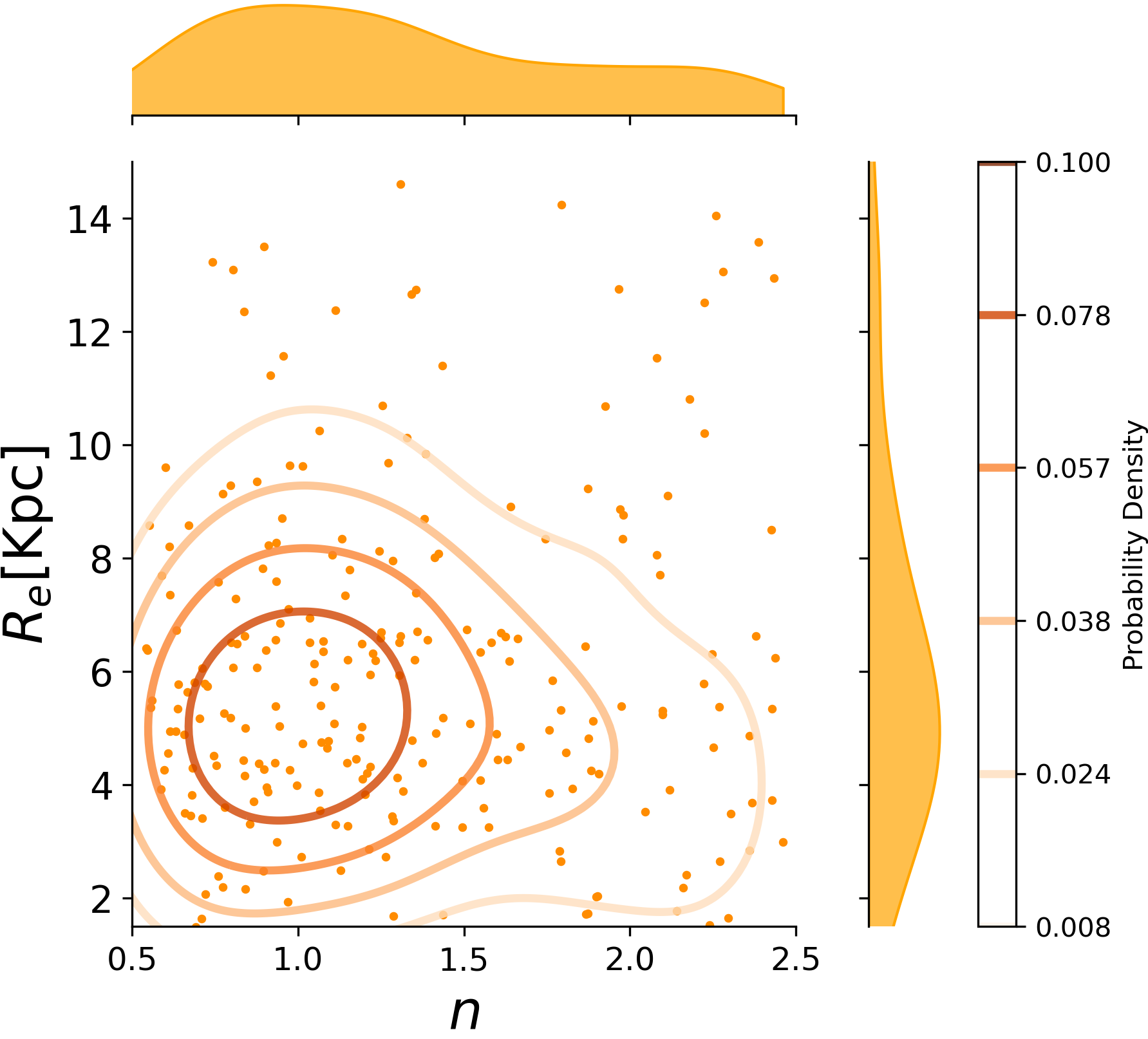}
    \includegraphics[scale=0.5]{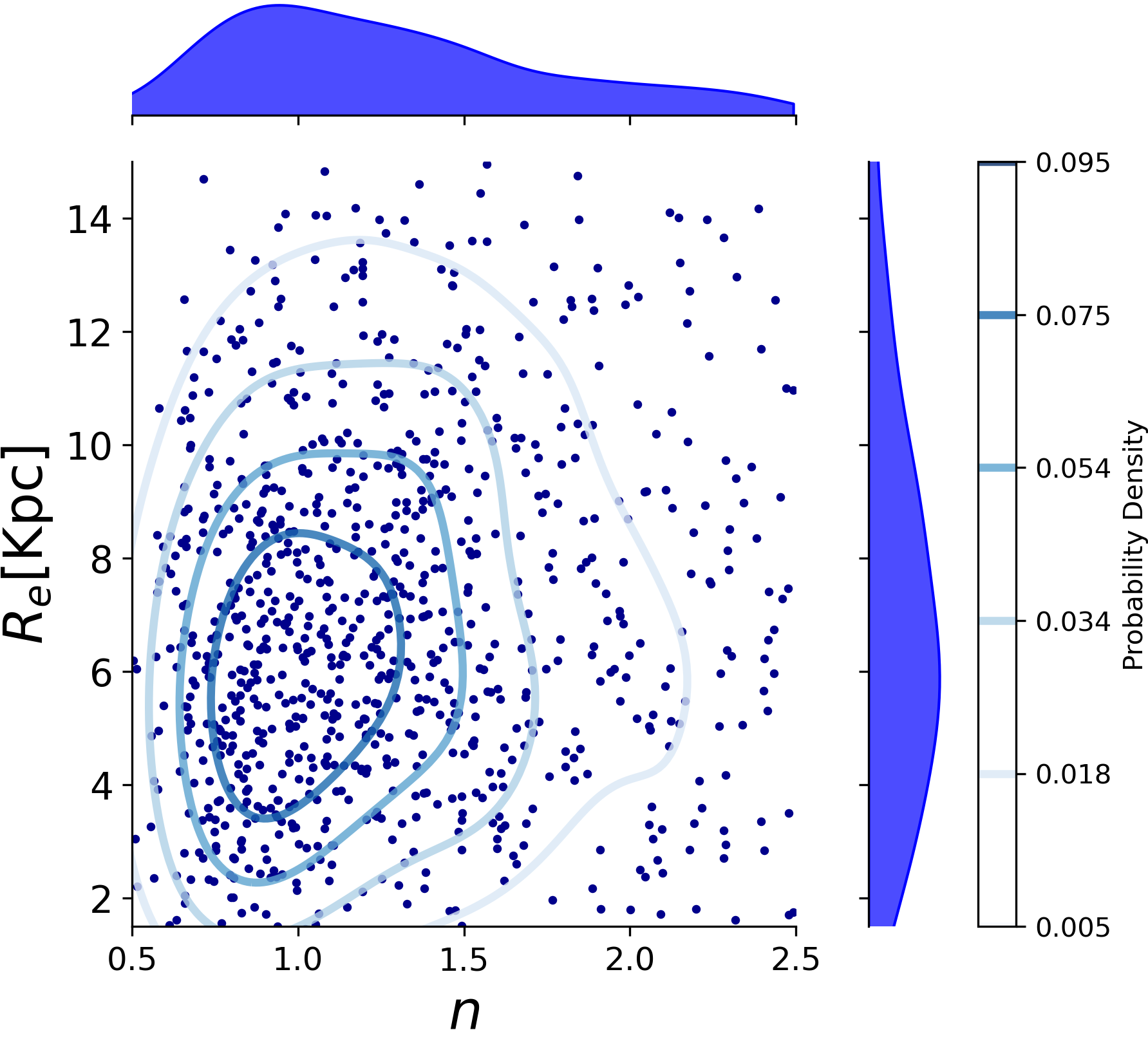}
    \caption{Effective radius–Sérsic index plane for LTGs in the $r$-band. Top panels show orange contours for the subsample divided by velocity dispersion bins, with the number of galaxies in each bin indicated in the upper left corner. Bottom panels display LTGs for all CGs (orange contours, left) and for the control sample (blue contours, right).}
    \label{fig:Re_n_sigma_late}
\end{figure*}

For Figure \ref{fig:Re_n_sigma_tran}, we used a kernel density estimator to visualise the data, with contour colours representing the joint probability density of $R_e-n$. In CGs with $\sigma_G < 200$ km/s, no bimodal distribution in $n$ is observed. However, in CGs with $200 < \sigma_G < 800$ km/s, a bimodal distribution becomes clear. The marginal plot reveals a higher fraction of galaxies with $n > 1.75$, accounting for 43\%, suggesting enhanced morphological transformation in these systems. At lower velocities, galaxies may transform slowly, while mergers could contribute to increasing the Sérsic index, though the merger rate is reported to be low \citep{1993Zepf}. A similar analysis for LTGs is presented in Figure \ref{fig:Re_n_sigma_late}, where $n$ remains unchanged regardless of the velocity dispersion bin.

\begin{table*}
\small % Reduce el tamaño de la fuente
\caption{Morphological type distribution cross CG velocity dispersion intervals}  
\centering
\begin{tabular}{|l|l|l|l|}
\hline

$\sigma_G$ intervals {[}$km/s${]} & Percentage of ETGs & Percentage of transition galaxies & Percentage of LTGs \\ \hline
0-200   (166)                        & $36.2\%$                    & $7.2\%$                                    & $30.1\%$                    \\ \hline
200-800 (149)                       & $47.2\%$                    & $7.1\%$                                   & $18.1\%$                    \\ \hline
\end{tabular}
\tablefoot{The percentages of ETGs, transition galaxies, and LTGs in the CG velocity dispersion intervals. The number of CGs per $\sigma_G$ interval is shown in the parenthesis.}
\label{tab:percentages}
\end{table*}

In Table \ref{tab:percentages}, we illustrate the percentage of each morphological type according to each sigma bin. The percentage of LTGs decreases in CGs with higher velocity dispersions, while the percentage of ETGs increases. For transition galaxies, the percentage is similar in both $\sigma_G$ ranges.

\subsection{The surrounding environment of CGs in our sample}
\label{sec:Surrounding}

We explore the environment of our CGs sample using the \cite{2007Yang} catalogue, following the methodology outlined by \cite{2021Zheng}. As mentioned in Section \ref{sec:data}, \cite{2007Yang} used a different approach to select the groups in their sample, focusing on selecting groups with different ranges of members rather than compact groups. Their method does not employ the isolation and compactness criteria, which are essential restrictions used by \cite{zheng2020compact} for identifying CGs.

We used catalogue III by \cite{2007Yang}, as they compiled three versions of the group catalogue using different sources of redshift data. Catalogue III includes all galaxies with spectroscopic information from SDSS, as well as data available at the time from other databases such as the 2dF Galaxy Redshift Survey  (\citealt{2001Colless}), the Point Source Catalog Redshift Survey (\citealt{2000Saunders}), and the Third Reference Catalogue (\citealt{1991DeVaucouleurs}). Additionally, it incorporates galaxies from SDSS that lack spectroscopic redshifts. For these galaxies, \cite{2007Yang} assigned the redshift based on their nearest neighbours.

We select all the groups and galaxies in the Stripe 82 that are included in the catalogue published by \cite{2007Yang} and perform a cross-match between the right ascension and declination positions of the galaxies in \citeauthor{2007Yang}'s groups and the galaxies in our sample of CGs. From this cross-match, we identify the IDs of the groups included in \cite{2007Yang} that had at least one galaxy that, according to our catalogue, is a galaxy in a CG. Using these group IDs, we selected all the galaxies that belong to the groups found by \cite{2007Yang}. We find that $278$ CGs that we have in our sample are identified as belonging to structures studied by \cite{2007Yang}, which corresponds to $88\%$ of our sample.

\cite{2007Yang} identified groups in the SDSS-DR7. As a consequence, many galaxies found in the catalogue published by \cite{zheng2020compact}, based on SDSS-DR14 and LAMOST data, are not included in the catalogue published by \cite{2007Yang}. Consequently, according to \cite{2021Zheng}, the remaining CGs are not found in the \cite{2007Yang} catalogue for two reasons:

First, the sky coverage of the galaxy catalogue used by \cite{zheng2020compact} is slightly larger than that of \cite{2007Yang}, as \cite{2007Yang} excluded galaxies located near the survey edges or in regions with very low completeness. This accounts for 26 CGs. The second reason is related to the faint-end magnitude limits in \cite{2007Yang}, which vary by position, ranging from 17.62 to 17.72 in extinction-corrected petrosian magnitude, due to different versions of target selection in the Main Galaxy Sample. In contrast, \cite{zheng2020compact} adopted a fixed magnitude limit of 17.77, this account for 12 CGs.

\begin{figure}
    \centering
    \includegraphics[scale=0.45]{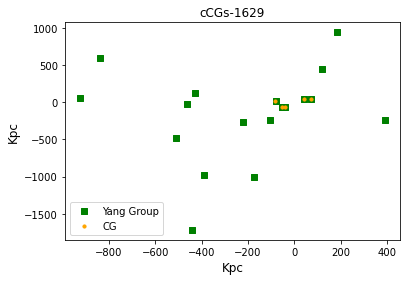}
    \includegraphics[scale=0.45]{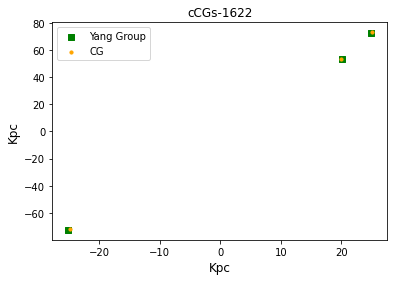}
    \includegraphics[scale=0.45]{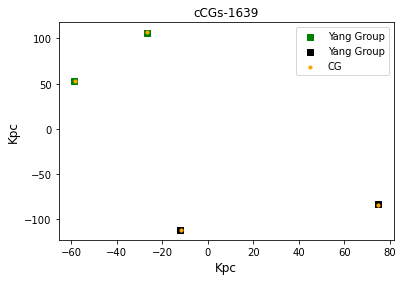}
    \caption{Visual representation of the spatial distribution and membership classification of CGs in different environments. The top panel shows a non-isolated CG (cCGs-1629), where the galaxies identified as part of the CG (orange points) are within a larger group (green squares), indicating their association with a major structure as reported by \cite{2007Yang}. The middle panel illustrates an isolated CG (cCGs-1622), where the galaxies correspond exactly with those identified by \cite{2007Yang}, confirming no additional members. In the bottom panel, a split CG is shown, where the galaxies in the CG (orange points) belong to different groups according to \cite{2007Yang}, specifically forming pairs (green and black squares), highlighting discrepancies in membership due to differing selection criteria.}
    \label{fig:ambiente_CG}
\end{figure}

From the 278 CGs, 122 CGs, which correspond to 44\%, are located within major structures, meaning they are subsets of larger groups and clusters. We will refer to these CGs as non-isolated from now on. This fraction of non-isolated CGs is what is also found in other studies (ranging from $20\%$ to $95\%$ \citealt{1998Barton}, \citealt{2005Andernach}, \citealt{2015Diaz}, \citealt{2021Zheng}). In the upper panel of Figure \ref{fig:ambiente_CG}, we present the projected distances in kpc on the sky of the galaxies within a CG, centred on the central coordinates of the CG. Here, we exemplify one of the non-isolated CGs, identified as cCGs-1629 by \cite{zheng2020compact}. The galaxies originally identified as part of the CG are represented by orange points. Additionally, the green squares represent galaxies in the counterpart group identified by \cite{2007Yang}, where we show that according to \cite{2007Yang}, the galaxies of cCGs-1629 form a subgroup within a larger group of 18 galaxies.

From the 278 CGs, $24\%$ of the CGs share the same members in both \cite{2007Yang} and our catalogue, which corresponds to 66. Adopting \cite{2021Zheng}'s terminology, we name these systems as isolated CGs. Indeed, there are no additional galaxies detected by Yang within their $r_{180}$ when compared to the galaxies in our catalogue of those CGs. In the middle panel of Figure \ref{fig:ambiente_CG}, we present an example of an isolated CG, identified by \cite{zheng2020compact} as cCGs-1622, where the three galaxies in this CG match the three galaxies of the group identified by \cite{2007Yang}.

The remaining $32\%$ of CGs, corresponding to 90 systems, includes 48 CGs where, according to the \cite{2007Yang} catalogue, their galaxies are associated with different groups, meaning they are not truly CGs. For example, in the bottom panel of Figure \ref{fig:ambiente_CG}, we show a CG with four galaxies, where, according to \cite{2007Yang}, the galaxies belong to two galaxy pairs represented in green and black. This difference is likely due to the velocity selection criteria applied by \cite{zheng2020compact}, which may overestimate the number of CGs. In fact, they refer to these groups as split CGs. The remaining 42 CGs contain fewer galaxies than in our sample, due to the magnitude limit in SDSS-DR7. As we cannot confirm that all these galaxies belong to the same CG, we decided not to include them in our analysis alongside the split CGs.

To summarise, we find that out of 278 CGs, at least one of the galaxies inside our sample of CGs was also identified by \cite{2007Yang}. Thus, excluding the 42 CGs that differ in sampling between \cite{2007Yang} and \cite{zheng2020compact}, those CGs where the groups in \cite{2007Yang} contain fewer galaxies due to the magnitude limit, we retain 236 CGs. Among these, 66 are classified as isolated CGs, accounting for 28\% of the 236 CGs, 122 (52\%) as non-isolated CGs, and 48 (20\%) as split CGs. These percentages are comparable to those reported by \cite{2021Zheng}, which are 28\%, 49\%, and 22\% for isolated, non-isolated, and split CGs, respectively. In the following two subsections, we consider only the galaxies in this 77\% of the 236 CGs, which corresponds to a total of 188 CGs.

\begin{figure}
    \centering
    \includegraphics[scale=0.4]{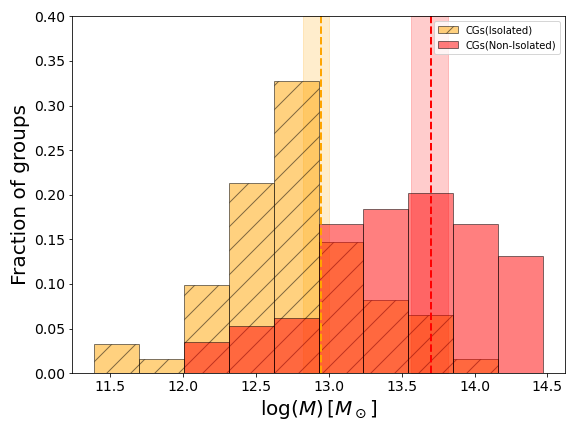}
    \includegraphics[scale=0.4]{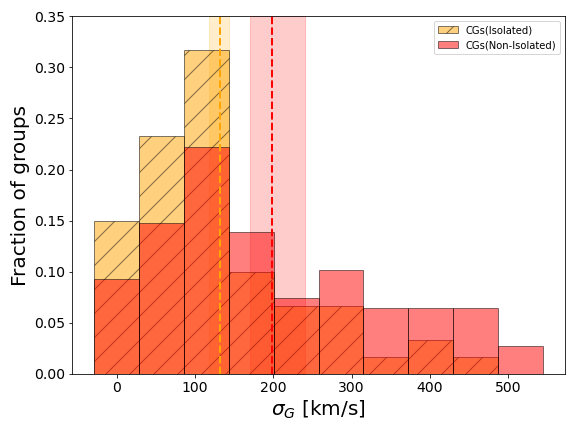}
    \caption{Histogram of the total stellar mass (top plot) of the groups investigated by \cite{2007Yang}, and the velocity dispersion of the CGs (bottom plot). In yellow we show isolated CGs, while in orange, the non-isolated ones. The median for each parameter in isolated and non-isolated CGs is represented by the dashed line, and the error is shown as the shaded region, which we estimate using bootstrapping with a 90\% confidence interval.}
    \label{fig:mases_I_E}
\end{figure}

In Figure \ref{fig:mases_I_E}, we show the fraction of groups where the CGs are located as a function of total stellar mass estimated by \cite{2007Yang} (top panel). In yellow, we show isolated CGs, and in orange, the non-isolated ones. Isolated CGs have lower mass distributions, while non-isolated CGs reach higher values. In fact, the median of the logarithm of the stellar mass of the groups where non-isolated CGs are found is $log(M_*[M_{\odot}])=13.7 \pm 0.2 $, with the error being the 90\% CI using bootstrapping, whereas for isolated CGs it is $log(M_*[M_{\odot}])=12.9 \pm 0.1$. This implies that group mass is crucial to understanding the galaxy's transformation. In the next section, we will discuss the impact of these masses on the morphological properties and the cessation of the SFR. In the bottom panel of Figure \ref{fig:mases_I_E} we present the fraction of CGs as a function of the velocity dispersion of each CG, following the same colour pattern. We find that isolated CGs typically have lower values of $\sigma_G$, with a median of $129.0\pm 14.8$ km/s, while non-isolated ones reach higher values, with a median of $187.6\pm 26.1$ km/s. 

It is important to consider this result because, as we have presented, there is a relationship between the morphology and dynamics of CGs (see Figure \ref{fig:sigma}). Additionally, \cite{2021Zheng} and \cite{2023Taverna}, found that the velocity dispersion of the CG increases with the density of the environment in which it is located, as it will be discussed in section \ref{sec:dynamics_struc}. This might suggest that these non-isolated CGs have recently fallen into these larger structures or are not dynamically individual subsystems but are more likely the result of chance alignments within larger systems. In the case of the first hypothesis \cite{2021Zheng} find that the embedded CGs in the outer parts of major structures are dynamically colder than the galaxies in the major structures. This suggests that these external CGs might, at least, consist of recently accreted groups. This is because it is expected that after the first pericentric passage, the group will disassemble and virialise, especially in structures like clusters (\citealt{Benavides2020}, \citealt{haggar2022}). In the second scenario, \cite{1986Mamon} and \cite{2021Zheng} suggest that CGs in the inner parts of major structures could be dominated by chance alignments. In addition, simulations show that between $30\%-60\%$ of CGs are chance alignments within larger groups (\citealt{2022Taverna}, \citealt{2020Hartsuiker}). These percentages vary depending on the assumed cosmology and the definition of CGs, as in some studies only those with four or more galaxies are considered, while in others also CGs with three members are considered. Additionally, \cite{Tzanavaris_2014} proposes that CGs are chance alignments in poor groups.

\subsection{Morphological transformation: Isolated versus non-isolated CGs}
\label{sec:tran_env}

\begin{figure}
    \centering
    \includegraphics[scale=0.5]{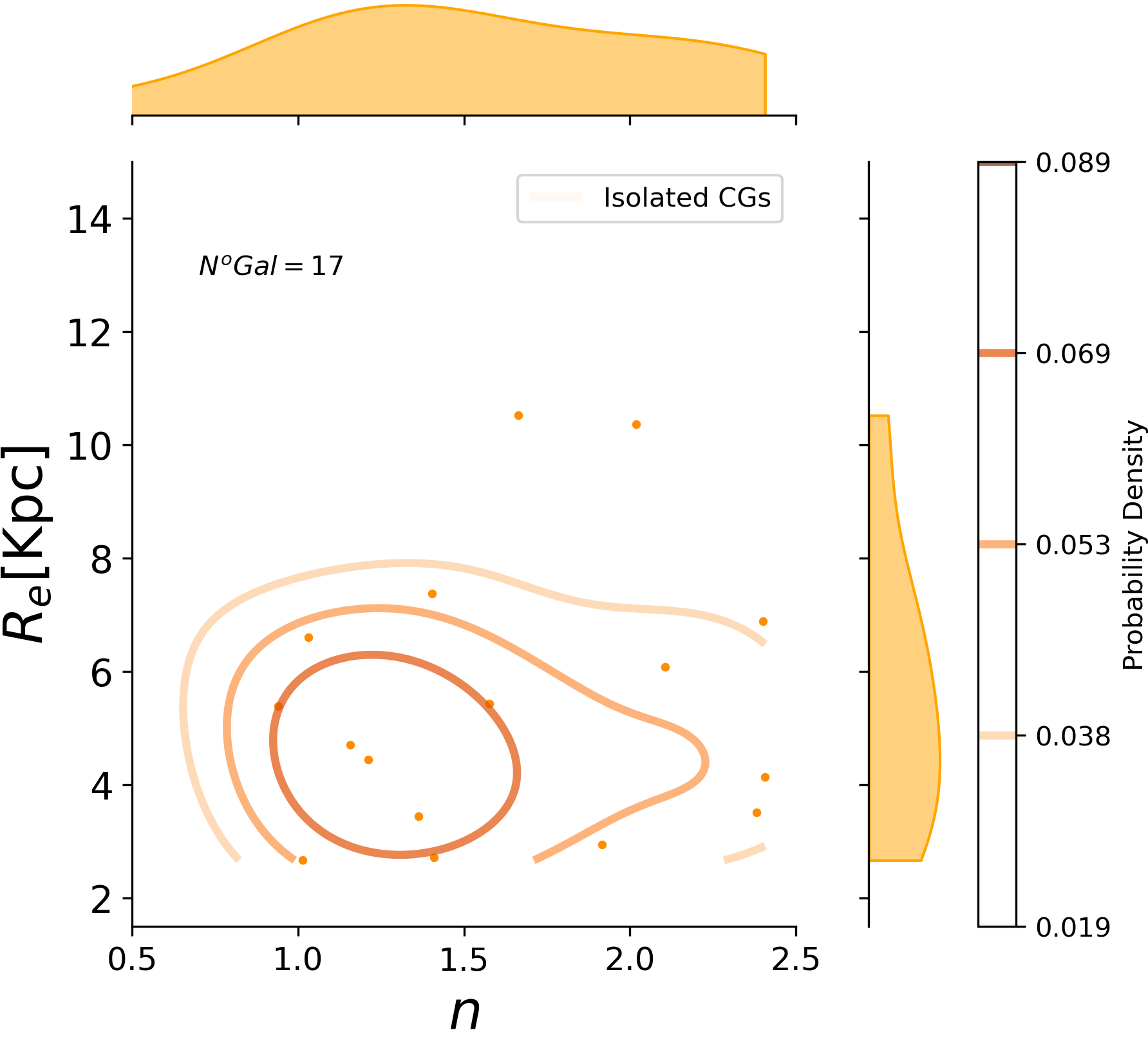}
    \includegraphics[scale=0.5]{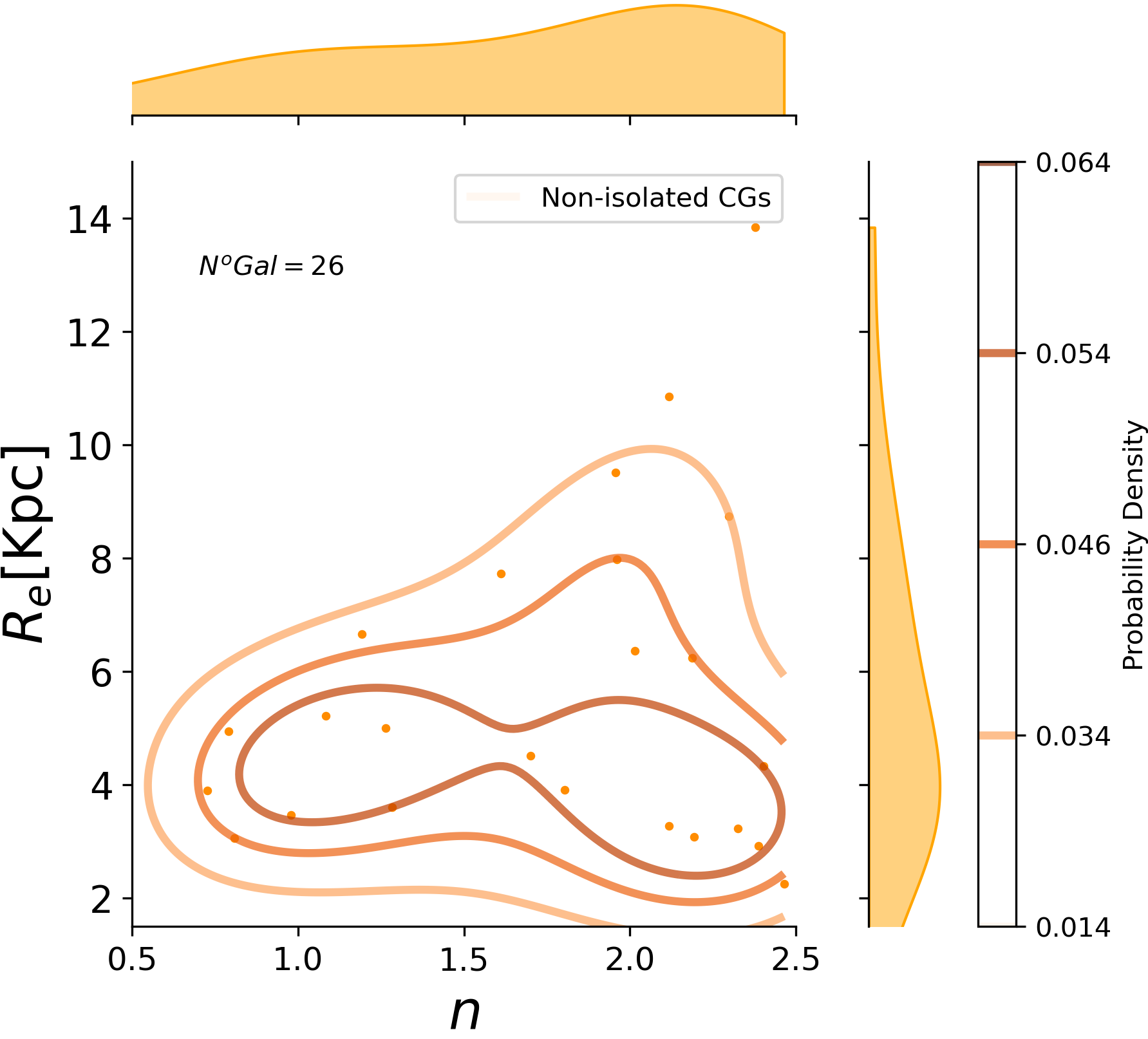}
    \caption{Contours of effective radius as a function of Sérsic index for transition galaxies, in $r$-band, in isolated CGs (top panel) and non-isolated CGs (bottom panel). The number of transition galaxies in each case is indicated in the upper right corner.}
    \label{fig:Re_n_I_E}
\end{figure}

Figure \ref{fig:Re_n_I_E} shows the effective radius as a function of the Sérsic index for transition galaxies in CGs in each environment (top panel for isolated CGs and bottom panel for non-isolated CGs). For transition galaxies in isolated CGs we do not find the bimodal distribution in the $R_e-n$ plane that we observed for all transition galaxies (see bottom left panel in Figure \ref{fig:Re_n_sigma_tran}). In fact, we observe that there is a high density in the $n-R_e$ plane for $n<1.75$. However, the contours for isolated CGs tend to stretch towards higher values of $n$ and lower values of $R_e$, where we find the population of peculiar galaxies within CGs.

In the case of the distribution of non-isolated CGs (see bottom panel of Figure \ref{fig:Re_n_I_E}), we observe the bimodality seen in all transition galaxies in the $R_e-n$ plane. In particular, when analysing the distribution of $n$, we find that $n$ increases smoothly until it reaches values characteristic of the peculiar galaxy population, i.e., $n > 1.5$. In fact, 62\% of the galaxies in our sample have $n > 1.5$. This finding suggests that the morphological transformation process appears to be more gradual in isolated CGs, whereas non-isolated CGs significantly populate the region of peculiar galaxies. This indicates that for most transition galaxies, the morphological transformation in non-isolated CGs has already occurred. In the top panel in Figure \ref{fig:mases_I_E}, we find that isolated CGs have lower mass distributions, while non-isolated CGs reach higher values, this could indicate that group mass is crucial for enhancing the morphological transformations observed in non-isolated CGs. Although it is important to note that Figure \ref{fig:Re_n_I_E} should not be directly compared with Figure \ref{fig:Re_n_sigma_tran}, because the first contains a subset of all the galaxies shown in Figure \ref{fig:Re_n_sigma_tran}.

\subsection{sSFR: Isolated and non-isolated CGs versus less dense environment}
\label{sec:sfr_env_cg}

In the previous section, we show that the environment in which the CGs are found enhances morphological transformation. Therefore, this leads naturally to the question of how much the major structure where the CGs are found can affect the other physical transformation compared with a less dense environment like the control sample, like sSFR. 
    
\begin{figure}
    \centering
    \includegraphics[scale=0.3]{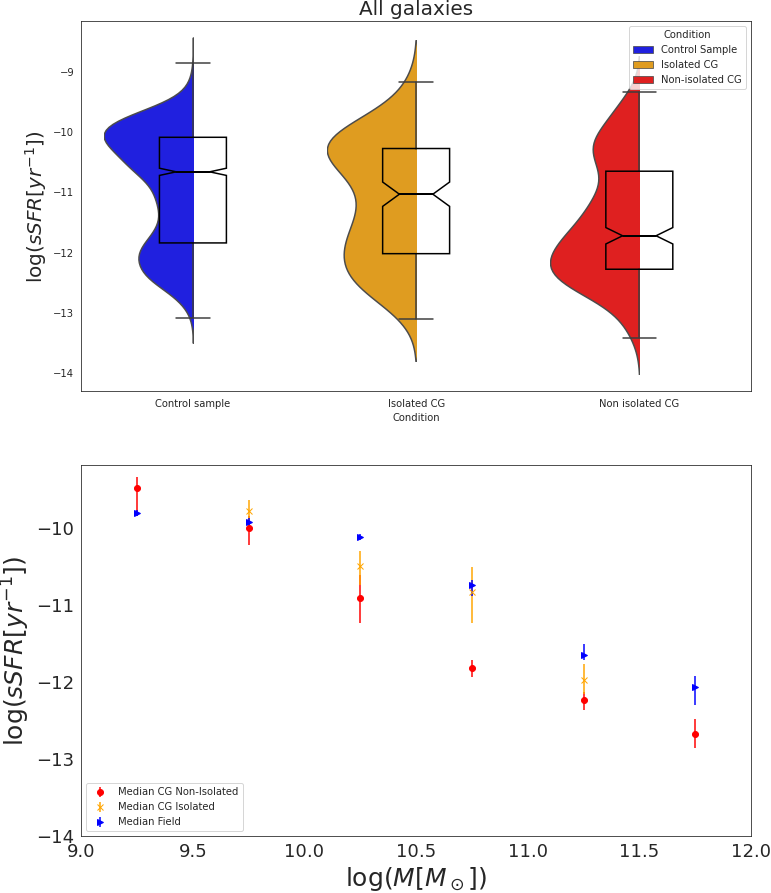}
    \caption{Median $log(sSFR)$ distribution for galaxies in different environments. The top panel shows box-plot and violent diagrams of the distribution of $log(sSFR)$ for galaxies in the different environments: control galaxies (blue), isolated CG galaxies (orange), and non-isolated CG galaxies (red). The bottom panel shows the median $log(sSFR)$ as a function of stellar mass, with error bars representing the 90\% confidence interval from bootstrapping.}
    \label{fig:sSFR_env}
\end{figure}

In Figure \ref{fig:sSFR_env}, the top panels show the box-plot and the violin diagrams of the distribution of log(sSFR[yr$^{-1}$]) of galaxies in different environments. Blue represents galaxies in the control sample, orange indicates isolated CG galaxies, and red shows non-isolated CG galaxies. The bottom panels show the median log(sSFR[yr$^{-1}$]) as a function of stellar mass, where we divide galaxies into stellar mass bins and calculate the median log(sSFR[yr$^{-1}$]) within each bin, where the points represent the median for galaxies in non-isolated CGs, the $x$ symbol represents the isolated CGs, and the triangles represent galaxies from the control sample. The error bars represent the 90\% confidence interval, estimated using bootstrapping.

In the top panel, the distribution of $log(sSFR)$ is compared between three samples of galaxies: Control sample, isolated CG, and non-isolated CG. In each box-plot diagram, the box represents the interquartile range of the $log(sSFR)$ distribution for each group, while the line inside the box shows the median value of $log(sSFR)$. The violin distribution presents the probability density distribution of $log(sSFR)$. Inspecting Figure \ref{fig:sSFR_env} we find that the control sample has a higher median, indicating these galaxies in this sample have more active star formation than galaxies in CGs, in fact, the notches of the three distributions do not overlap, indicating statistically significant differences in the distribution (\cite{Mcgill01021978}, \cite{krzywinski2014comparing}) of $log(sSFR)$ among the three analysed environments. Therefore, the galaxies in isolated and non-isolated CGs have lower median $log(sSFR)$, suggesting that CGs can inhibit star formation, with the environment in which the CGs are located affecting the star formation of their members.
    
Additionally, the $log(sSFR)$ values show greater dispersion in isolated CGs, indicating more variability in star formation activity compared to non-isolated CGs, which have a more compact distribution. This behaviour in the interquartile range can be seen in the violin distribution, where a clear bimodality is observed for both the control sample and isolated CGs. However, in the case of non-isolated CGs, this bimodality is less pronounced, with a peak at $log(sSFR[yr^{-1}])\sim-12.2$. This suggests that interactions between CGs and their surroundings may contribute to greater suppression of star formation.
    
The bottom panel shows how the specific star formation rate varies with stellar mass. Galaxies in CGs, both isolated and non-isolated, have lower $log(sSFR)$ values at intermediate stellar masses ($10.0<M(M_{\odot}])<11.5$) compared to the control sample.

\begin{figure*}
    \includegraphics[scale=0.222]{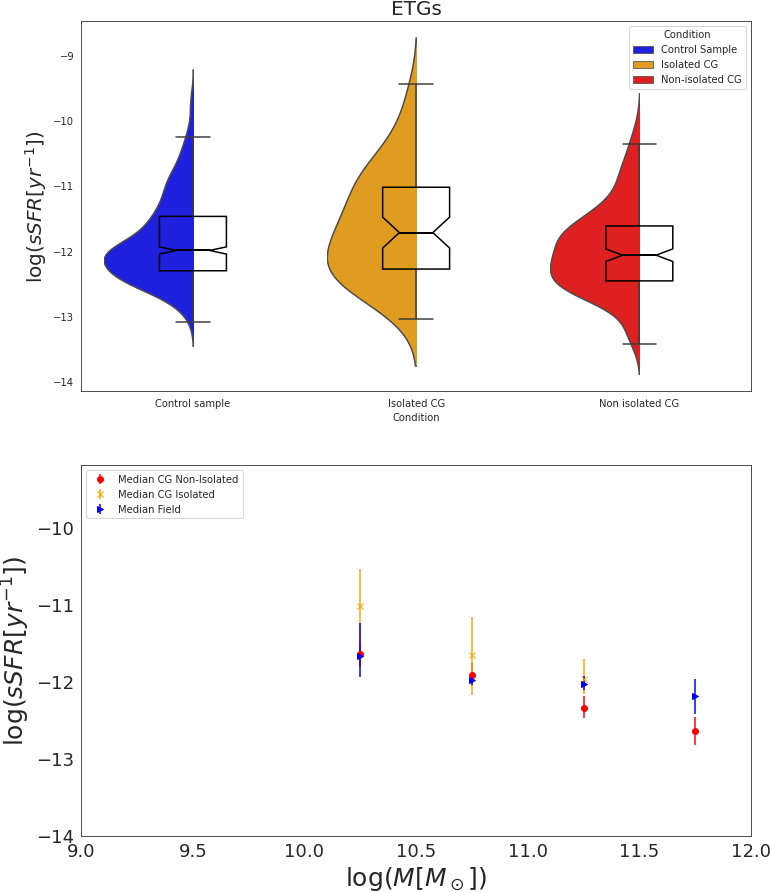}
    \includegraphics[scale=0.222]{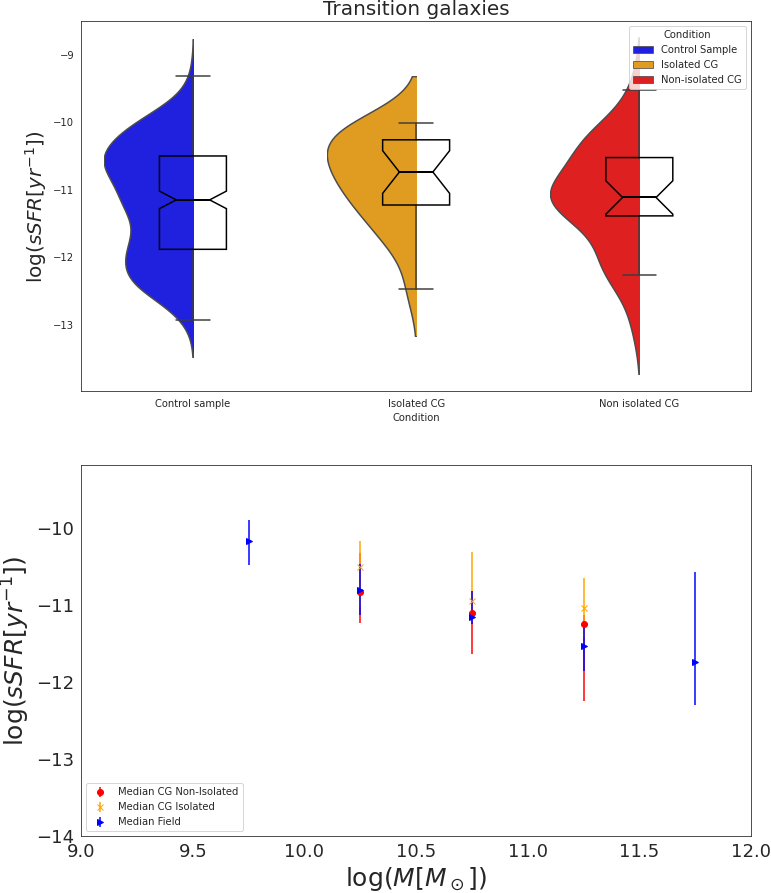}
    \includegraphics[scale=0.222]{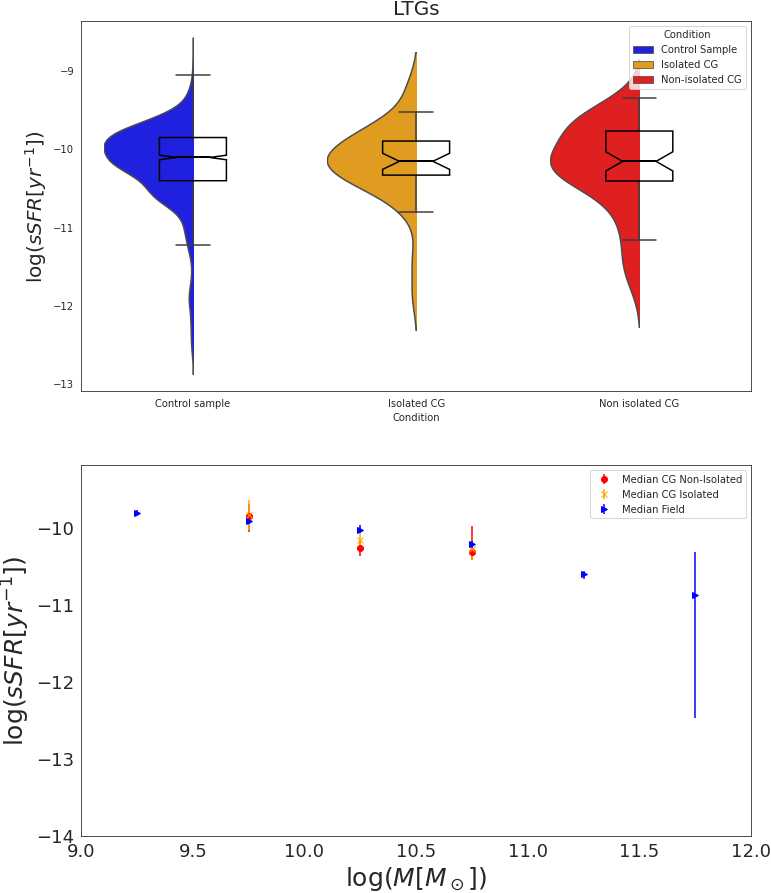}
    \caption{Same as Figure \ref{fig:sSFR_env}, but for ETGs (left panel), transition galaxies (middle panel), and LTGs (right panel).}
    \label{fig:sSFR_env_ETG}
\end{figure*}

In \cite{Montaguth2023}, we showed that there are statistically significant differences in the $log(sSFR)$ distribution for LTGs and ETGs, but not for transition galaxies. Figure \ref{fig:sSFR_env_ETG} shows the results for different galaxy morphologies: ETGs (left panel), transition galaxies (middle panel), and LTGs (right panel), in the same way as in Figure \ref{fig:sSFR_env}.
    
In the top left panel in Figure \ref{fig:sSFR_env_ETG} we show the median $log(sSFR)$ is higher for ETGs in isolated CGs compared to their counterparts in the control sample and non-isolated CGs. Additionally, the interquartile range, which represents the range between the 25th and 75th percentiles of the data, is wider for isolated CGs than for the other two environments. The whiskers also extend to higher $log(sSFR)$ values, indicating a greater dispersion towards higher $log(sSFR)$ compared to the other environments. In contrast, for ETGs in non-isolated CGs, the median $log(sSFR)$ is lower than in the other environments, and its interquartile range does not extend as far as that of isolated CGs. This behaviour is also reflected in the violin distribution, where the distribution of ETGs in isolated CGs exhibits higher values, which is not observed in non-isolated CGs.
    
To analyse how $log(sSFR)$ varies with stellar mass across different environments (see left bottom panel in Figure \ref{fig:sSFR_env_ETG}), we focus on two trends. First, in the lowest stellar mass bins, ETGs in isolated CGs show higher $log(sSFR)$ compared to those in the control sample, although the overlap of the error bars suggests that this difference is not statistically significant. In contrast, at higher stellar masses, ETGs in non-isolated CGs exhibit lower $log(sSFR)$ than their counterparts in the control sample. These findings suggest that isolated CGs help sustain star formation in low-mass ETGs, while non-isolated CGs promote the cessation of star formation in high-mass ETGs.

For transition galaxies, as shown in the middle top panel of Figure \ref{fig:sSFR_env_ETG}, the median $log(sSFR)$ is higher in isolated CGs compared to both the control sample and non-isolated CGs. In non-isolated CGs, the median $log(sSFR)$ is lower than in isolated CGs but similar to the control sample. However, the interquartile range for non-isolated CGs is smaller, indicating that the $log(sSFR)$ values are more concentrated and exhibit lower dispersion compared to the control sample. The larger interquartile range in the control sample reflects a higher dispersion, likely due to the bimodal distribution of $log(sSFR)$ observed in the violin distribution.

The distribution of $log(sSFR)$ further underscores these environmental differences. In isolated CGs, transition galaxies peak at $log(sSFR)$ values corresponding to the high-$log(sSFR)$ peak of the control sample’s bimodal distribution. In contrast, transition galaxies in non-isolated CGs peak near the middle of the bimodality seen in the control sample. Notably, neither isolated nor non-isolated CGs show a clear bimodality in their $log(sSFR)$ distributions. These observations suggest that the environment of CGs plays a crucial role in regulating and, ultimately, ceasing star formation in transition galaxies.
    
Looking at how $log(sSFR)$ behaves as a function of stellar mass across different environments (see the middle bottom panel of Figure \ref{fig:sSFR_env_ETG}), we observe no significant differences between isolated and non-isolated CGs compared to the control sample. The differences are marginal and within the error bars.

For LTGs, in the right top panel in Figure \ref{fig:sSFR_env_ETG}, the median is similar for both isolated and non-isolated CGs and the control sample. In the case of LTGs in non-isolated CGs, we see that the interquartile range is larger compared to isolated CGs and the control sample. This indicates that the distribution of $log(sSFR)$ for LTGs in non-isolated CGs has a larger dispersion. This is consistent with the violin distribution. When comparing how $log(sSFR)$ behaves as a function of stellar mass, we observe no significant differences between isolated and non-isolated CGs and the control sample. The largest difference is in the lowest stellar mass bin, where non-isolated CGs have a higher median $log(sSFR)$ than the control sample.
    
However, when dividing the galaxies by morphology, we observe that the differences between the control sample and the galaxies in isolated and non-isolated CGs are not statistically significant, as the notches of the boxes overlap. In Table \ref{tab:quenched_galaxies}, we present the percentage of quenched galaxies in isolated and non-isolated CGs, as well as in the control sample. We consider a galaxy to be quenched if $log(sSFR)\leq-11$, based on the criterion proposed by \cite{2013Wetzel}.
    
It is important to emphasise that this threshold is close to the value that defines the bimodality observed in transition galaxies of the control sample (see middle panel in Figure \ref{fig:sSFR_env_ETG}), which is $log(sSFR)=-11.46$. This value was estimated by fitting two Gaussians and determining their intersection point. While this could serve as a definition for when a galaxy is quenched, we selected the \cite{2013Wetzel} criterion to ensure comparability with other studies.
    
Observing the first row of the Table \ref{tab:quenched_galaxies} for all galaxies, we see that the fraction of quenched galaxies increases from the control sample to the non-isolated CGs. When analysing the three morphological types, we find that non-isolated CGs have a higher percentage of quenched galaxies for ETGs, transition galaxies, and LTGs compared to both isolated CGs and the control sample. Simulations suggest that galaxies are more likely to enter a quenched state in structures with masses between $10^{13.5}$ and $10^{14.5} M_{\odot}$ (\citealt{2019MNRAS.488..847P}). This mass range is similar to the peak of the mass distribution of non-isolated CGs, as shown in Figure \ref{fig:mases_I_E}. In contrast, for isolated CGs, we find that the fraction of quenched galaxies is lower than in the control sample for ETGs and transition galaxies.

\begin{table*}
\small % Reduce el tamaño de la fuente
\centering
\caption{Fraction of quenched galaxies across morphologies in different environments}
\renewcommand{\arraystretch}{1.5} % Aumenta el tamaño de las filas
\begin{adjustbox}{max width=\textwidth}
\begin{tabular}{|c|ccc|}
\hline
Galaxy morphology   & \multicolumn{1}{c|}{Control sample} & \multicolumn{1}{c|}{Isolated CGs} & Non-isolated CGs \\ \hline
All galaxies        & \multicolumn{1}{c|}{$0.42\pm0.01$}           & \multicolumn{1}{c|}{$0.52\pm0.04$}         & $0.71\pm0.02$             \\ \hline
ETGs                & \multicolumn{1}{c|}{$0.87\pm0.02$}           & \multicolumn{1}{c|}{$0.77\pm0.05$}         & $0.91\pm0.02$             \\ \hline
Transition galaxies & \multicolumn{1}{c|}{$0.53\pm0.03$}           & \multicolumn{1}{c|}{$0.39\pm0.10$}         & $0.67\pm0.08$             \\ \hline
LTGs               & \multicolumn{1}{c|}{$0.06\pm0.01$}           & \multicolumn{1}{c|}{$0.05\pm0.03$}         & $0.12\pm0.04$             \\ \hline
\end{tabular}
\end{adjustbox}
\tablefoot{Fraction of quenched galaxies for different galaxy morphologies in the control sample, isolated CGs, and non-isolated CGs. The criterion proposed by \cite{2013Wetzel} ($log(sSFR)\leq -11$) is used to determine if a galaxy is quenched.}
\label{tab:quenched_galaxies}
\end{table*}

\section{Discussion}
\label{sec:discussion}

In this study, we examine the morphological and physical characteristics ($R_e$ and $n$) using S-PLUS images and $sSFR$ of galaxies within a sample of CGs, selected using the catalogue \cite{zheng2020compact}. This sample is then subdivided into isolated and non-isolated CGs, following the analysis done by \cite{2007Yang}. We explore how these properties change based on the environment in which the CGs are situated. Our findings reveal significant morphological differences in transition galaxies between isolated and non-isolated CGs. Additionally, non-isolated CGs show a higher fraction of quenched galaxies compared to their isolated counterparts and the control sample. In the following section, we delve deeper into our results, comparing them with other studies, and we propose an evolutionary scenario based on these observational outcomes.

\subsection{Can the environment of the CGs influence a physical transformation?}

In \cite{Montaguth2023} we compared the same sample of CGs analysed in this work with a control field sample, and suggested that environmental effects lead to a higher proportion of quenched galaxies and a lower median $log(sSFR)$ in CGs. This indicates a cessation of star formation, regardless of galaxy type, where a possible explanation for these differences could be tidal interactions, shocks, and turbulence (\citealt{2015bAlatalo}, \citealt{2016Bitsakis}), for the specific sample that we studied in that work. In this work, we find that both isolated CGs and non-isolated CGs have higher fractions of quenched galaxies compared to the control sample, and that this fraction is higher for non-isolated CGs with respect to the isolated counterpart. 

For isolated CGs, the mechanism that likely contributes to their higher fraction of quenched galaxies compared to the field is tidal interactions, which can produce neutral gas loss and heating, enriching the intra-group medium.  In the case of non-isolated CGs a possible scenario is that galaxies may experience different mechanisms depending on the mass of the group they belong to, resulting in this higher percentage of quenched galaxies. For non-isolated CGs located in low-mass structures, galaxy interactions are expected to be the main driver of transformations. However, for non-isolated CGs located in more massive structures, mechanisms such as ram-pressure stripping (\citealt{2021Roberts}) and galaxy harassment (\citealt{1996Moore}) may play a role.

These non-isolated CGs exhibit higher velocity dispersions (see the bottom panel in Figure \ref{fig:mases_I_E}), resulting in smaller crossing times, which is indicative of a more dynamically evolved state compared to the isolated CGs. This dynamical difference allows us to observe a lower $log(sSFR)$ and a larger fraction of quenched galaxies in non-isolated CG. Thus, the physical properties of the CG galaxies are affected by interactions between galaxies within the CG, and the interaction of the CG with its surrounding environment.

However, when dividing galaxies by morphological type, the differences of the distribution of $log(sSFR)$ between environments tend to diminish. This uniformity within each morphological type reduces the ability to detect significant environmental differences. Additionally, the smaller sample sizes when splitting by morphology can reduce the statistical significance of these effects. There is also a clear relationship between morphology, star formation, and environment. In denser environments, a higher fraction of quenched galaxies and ETGs is expected (\citealt{1984Dressler}, \citealt{2003Kauffmann}), where most ETGs are expected to be quenched (\citealt{valentina2015}). This could explain why we do not find major differences between the control sample and the isolated and non-isolated CGs.

Furthermore, \cite{2023Perez-millan} found that, at fixed morphology, there are no significant differences in the $SFR-M_{*}$ relation between cluster and field galaxies. However, when analysing the star formation histories (SFHs) of cluster galaxies, they display faster quenching than those of field galaxies at fixed masses and morphologies. This indicates greater quenching efficiencies due to external agents. In this sense, the SFH provides a comprehensive view of how a galaxy has formed stars throughout its lifetime, while the $SFR$ only reflects the star formation rate at a specific moment in time. 

Interestingly, we found differences in the $sSFR$ for ETGs. In the high stellar mass bins, the sSFR is lower in non-isolated CGs compared to the control sample, indicating an environmental effect. Furthermore, \cite{2020Moura} shows that ETGs in high-$\sigma$ CGs are older, more metal-rich, and have undergone more efficient quenching of star formation than ETGs in low-$\sigma$ CGs. This aligns with the idea that high-$\sigma$ CGs are more dynamically evolved than low-$\sigma$ CGs. In our case, non-isolated CGs have a higher median velocity dispersion.

In the case of the low stellar mass bins, we found that the median $sSFR$ for all the galaxies is higher in isolated CGs compared to the control sample. This could suggest that galaxies may experience episodes of stellar rejuvenation through the accretion of gas from their surroundings, which reactivates star formation, particularly in groups (\citealp{2007Rampazzo}). This underscores the necessity of analysing the star formation history to gain a better understanding of the processes occurring in these systems in a subsequent analysis.

\subsection{Evolutionary scheme: the role of the dynamics in CGs and its connection with major structures}
\label{sec:dynamics_struc}

As presented in Section \ref{sec:results}, there is a clear relationship between the dynamics of CGs and morphological transformation. Additionally, we find that approximately 41\% of all CGs are not isolated, this represents a lower limit due to the lack of information about the surrounding environment for 45 CGs in our study. The surrounding environment of these non-isolated CGs influences their observed morphological transformation in various ways. Specifically, these CGs exhibit lower sSFR compared to their isolated counterparts, for fixed masses in the range of $10.0<log(M_*/M_{\odot})<11.5$.

Furthermore, according to \cite{2021Zheng}, CGs embedded in larger structures show a correlation between their velocity dispersion and the values displayed by their parent structure in which they are embedded (i.e., a larger group or cluster). This relationship follows an almost one-to-one pattern for CGs with velocity dispersions lower than 500 km/s, such that as the velocity dispersion of the structures that contain the CGs increases, the velocity dispersion of the CGs also increases. We find that the non-isolated CGs in our sample have a higher median velocity dispersion than isolated CGs.

Therefore, by considering the velocity dispersion of CGs, we can speculate whether these systems are part of larger structures. In any case, a detailed spectroscopic analysis of the CGs’ environment is required to fully understand their connection with larger-scale structures.

It is important to note that what we refer to as non-isolated CGs are, in fact, a mixture of embedded and predominant CGs as defined by \cite{2021Zheng}. The embedded CGs are found in major structures, with the luminosity of the CG being lower than that of the other galaxies in the major structure. In contrast, predominant CGs have higher luminosity than the additional galaxies in the major structure. The correlation between the velocity dispersion of the parent group and the CGs found by \cite{2021Zheng} applies only to embedded CGs, as they did not explore the case of predominant CGs. Recently, \cite{2023Taverna} studied the properties of 1368 CGs catalogued by \cite{2022Zandivarez} using SDSS-DR16. They found that the velocity dispersion of CGs increases with the density of the environment in which they are located.

\begin{figure*}
    \centering
    \includegraphics[scale=0.4]{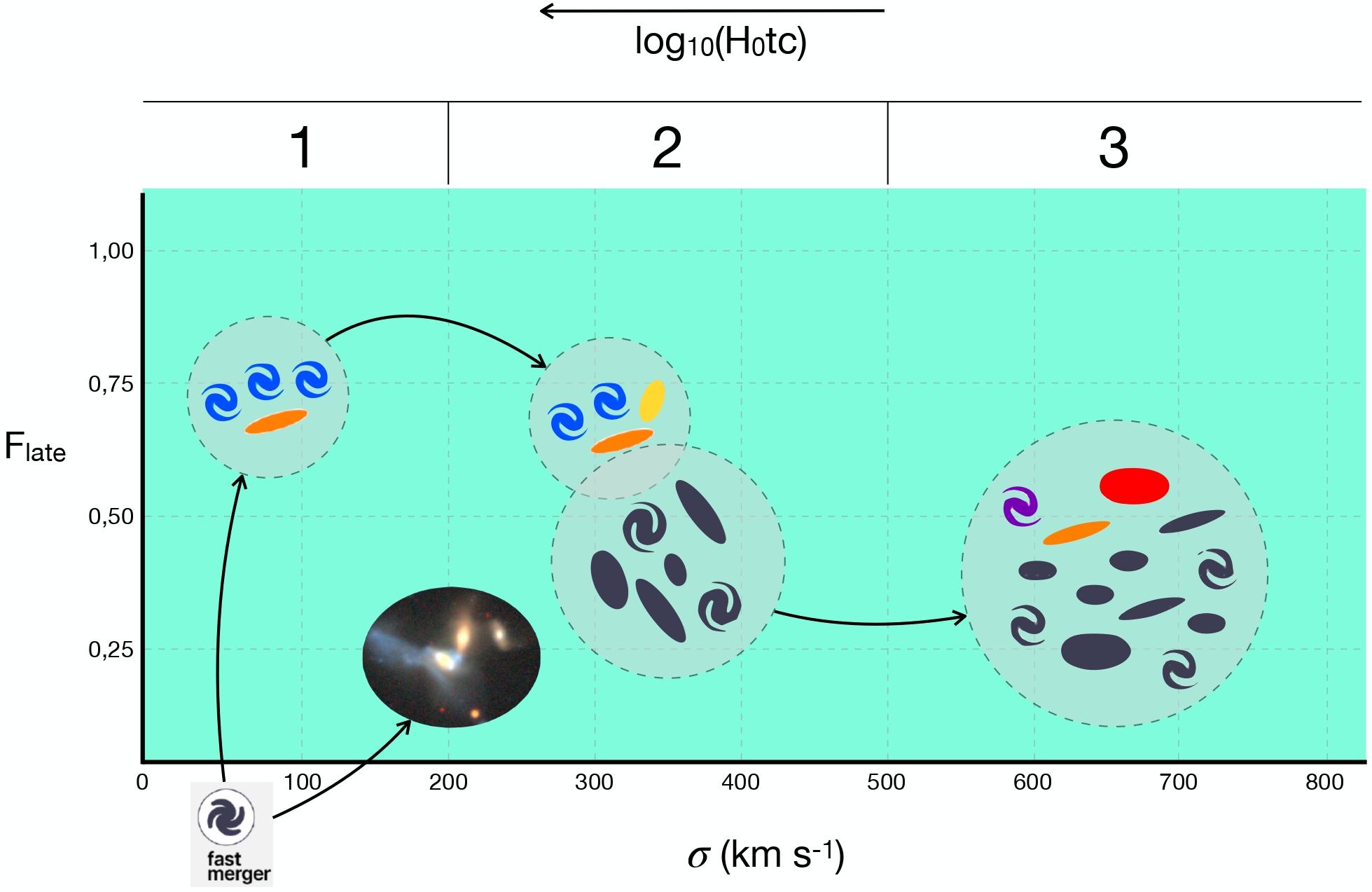}
    \caption{Evolutionary scheme we propose for CGs. In the x-axis we have the velocity dispersion of CG, in the opposite direction the crossing time which increases as $\sigma_G$ decreases, having less dynamically evolved CGs. On the y-axis, we have the fraction of LTGs. Our scheme evolves from stages 1 to 3, where the CGs have a lower fraction of LTGs, higher $\sigma$, and lower $H_{o}t_{c}$. Our results show that this evolution is accelerated in non-isolated CGs.}
    \label{fig:esquema}
\end{figure*}

Combining the results shown in Section \ref{sec:Disperson_mor}, we can further categorise CGs into three distinct stages, allowing us to propose an evolutionary scheme, based on the morphological characteristics of their members and the environments they inhabit. These stages allow us additionally to establish correlations with the velocity dispersion of the systems, as mentioned in Section \ref{sec:Surrounding}. High $\sigma_G$ values in this context may indicate a stronger association between the dynamics of CGs and the larger-scale structures they are embedded in. In Figure \ref{fig:esquema}, we represent how these three stages correlate. In the first stage, there are CGs rich in LTGs, with larger crossing times, which are in an initial or less dynamically evolved state and isolated. In this stage the transition galaxies start to become smaller and more compact, appearing as a peculiar galaxy population in CGs, with respect to the control sample, in the $R_e-n$ diagram. Tidal forces pull matter out of the galaxies, which increases the amount of neutral gas in the intra-group medium. 

This HI gaseous component medium has been observed by several authors (\citealt{2001Verdes-Montenegro}, \citealt{2022Jones}, \citealt{2023Cheng}). The second stage includes CGs with a lower fraction of LTGs and smaller crossing times. Here the environment in which the CGs are located plays a relevant role in the evolution of the galaxies contributing to the transformation process that we observe in the transition galaxies, increasing the population of galaxies that characterises the CGs, that is the `peculiar' population of transition galaxies. In this stage, there is a combination of tidal effects produced by galaxy-galaxy interactions inside the CGs, and the interaction of the CG with its surrounding environment. Finally, in the last stage probably most of the transition galaxies have already suffered the morphological transformation that we observe in systems with lower $\sigma_G$. It is possible that during this stage, CGs may begin to disrupt within the major structures they inhabit.

Therefore, we suggest that compact groups (CGs) can evolve from scenarios 1 to 3. CGs in stage 1 have three potential ways of evolving. The first is that these CGs will merge due to dynamical friction (\citealt{1992Mamon}, \citealt{2008Jiang}), since the merger rate in groups is inversely related to the $\sigma_G^3$ of the group/cluster.

The second possibility is that other groups can accrete them, or they can accrete other galaxies and reach stage 2. The third option is for the CGs to remain virialised, depending on the mass of the halo containing them (\citealt{2001GomezFle}).

In stages 2 and 3, the CGs and their environment can be accreted by major structures, which accelerates and promotes the morphological transformation of the transition galaxies. Hence, the CGs would move from top to bottom in our scheme presented in Figure \ref{fig:esquema}, decreasing the fraction of late-type galaxies (LTGs), and from left to right, increasing their velocity dispersion.

The increase in velocity dispersion may be related to a growing mass of the CGs due to the accretion of galaxies or because the CGs are being accreted by other groups. In the first case, this would probably only be seen as a slight increase in velocity dispersion, while in the second case, the dynamics of the CGs would be dominated by the larger structure in which they are embedded.

In both scenarios, we would observe the morphological transformation of the galaxies within these groups being accelerated, along with low values for the crossing time. According to simulations, when a group falls into structures like clusters, the group becomes dynamically hotter, even though the relative velocity of the group remains, on average, lower than that of the galaxies in the cluster.

Additionally, it is expected that after the first pericentric passage, the group will disintegrate (\citealt{2019Bahe}, \citealt{Benavides2020}, \citealt{haggar2022}). Our proposed evolutionary scenario aligns with \cite{2004Coziol}, who suggest that the formation of CGs embedded in massive structures occurred earlier than the formation of isolated CGs, a viewpoint also supported by \cite{2006Tovmassian}.

However, we note that the study conducted by \cite{2004Coziol} was based on a small sample of 27 Hickson CGs, where only $\sim 30\%$ of CGs were isolated, while the remaining 70\% were part of major structures (groups and clusters).

Our results (sections \ref{sec:tran_env} and \ref{sec:sfr_env_cg}) support the idea that CGs could be responsible for the morphological transformation of galaxies, which is enhanced by the environment in which they are located. They are also responsible for quenching the star formation of galaxies, where the environment again plays a relevant role, by enhancing this process. An extreme example that supports this scenario is the CG falling into the Abell 1367 cluster (Blue infalling group, \citealt{2006Cortese}). Indeed, \cite{2006Cortese} showed that due to the combination of the effect of the tidal forces in the CG, combined with the ram-pressure of the intra-cluster medium, the galaxies of the CG look fragmented, and the ionised gas is out off galaxies and ejected, generating the HII regions that are observed close to the galaxies of the CG. Hence, the evolution of galaxies in CGs is affected both by the local environment, i.e., by the fact of being in CGs, and also by the fact that these CGs are part of major structures. This should be taken into account when analysing CGs, given that there is a complex interplay of physical phenomena occurring at both local and major scales.

\section{Summary and conclusions}
\label{sec:conclusions}

This work is the second paper of a series aimed at understanding the evolution of galaxies in compact groups (CGs). In this study, we examine 316 CGs located in the Stripe 82 region, selected from the catalogue published by \cite{zheng2020compact}, using images in 12 bands from the S-PLUS project. We subdivide our sample of CGs into isolated and non-isolated, following a previous study done by \cite{2021Zheng} and by using the catalogue published by \cite{2007Yang}. Our goal is to understand the dynamics of CGs as well as how the environment surrounding them impacts the morphological and physical properties of their galaxy members. We compare our results against a sample of field galaxies, in order to determine the differences within a less dense environment. Galaxies are classified into early-type (ETG), late-type (LTG), and transition galaxies based on their Sérsic index and colour. In the following, we summarise our main findings: 

\begin{enumerate}
    
    \item We find that CGs with higher velocity dispersions, higher compactness, higher ETG fraction, and lower crossing times, are dynamically more evolved than those CGs with shorter velocity dispersion, a lower ETG fraction, and larger crossing times.

    \item CGs with higher velocity dispersions exhibit a clear bimodal distribution of their transition galaxies in the $R_e-n$ plane. However, this bimodality is not clear for transition galaxies located in CGs having lower velocity dispersion CGs. This suggests an enhanced morphological transformation process in higher velocity dispersion CGs, where peculiar galaxy populations prevail. Conversely, lower velocity dispersion CGs may indicate a slower morphological transformation, possibly due to gradual changes thanks to the tidal interactions. Although we cannot rule out the scenario that mergers within the CG might also contribute to the morphological transformation. This suggests that the CG environment is ideal for studying the morphological changes that galaxies undergo.
    
    \item We find that at least $39\%$ of our 316 CGs are non-isolated, i.e., they are in major structures.
    
    \item Non-isolated CGs have a wider range of velocity dispersions, reaching $\sigma_G < 800 km/s$. These non-isolated CGs have higher median velocities dispersion than isolated CGs. Isolated CGs have a narrower range, reaching $\sigma_G \leq 400 km/s$. Therefore, the value of the velocity dispersion of a CG may indicate whether the CG is associated with a larger structure.
    
    \item We identify that the transition galaxies in isolated or non-isolated CGs do not follow the same bimodal effective radios and Sérsic index ($R_e-n$) distribution as all transition galaxies, but they contribute to one of the two parts of this bimodality. Indeed, for transition galaxies in isolated CGs, we find a high density in the $R_e-n$ plane for $n<1.75$, suggesting that most of these galaxies have not yet undergone morphological transformation. In contrast, In transition galaxies in non-isolated CGs, we find a bimodality; specifically, when we analyse the $n$ distribution, it increases smoothly towards higher values, with 62\% of the galaxies in this sample having $n > 1.5$, reaching the $n$ values characteristic of the peculiar galaxy region. This suggests that a high percentage of these galaxies have already undergone a morphological transformation. These findings indicate that the large-scale environment where CGs reside enhances morphological transformation. 
    
    \item We find that galaxies in CGs exhibit a lower mean sSFR compared to the control sample, suggesting that environmental density may inhibit star formation. In particular, galaxies in non-isolated CGs have even lower sSFR rates than those in isolated CGs, indicating that interactions within these dense environments may lead to greater suppression of star formation. Furthermore, a higher fraction of quenched galaxies is observed in non-isolated CGs compared to those in isolated CGs and the control sample.
    
    \item We compare the sSFR at different morphological types and find significant differences only for ETGs. Those in isolated CGs exhibit a lower fraction of quenched galaxies compared to the control sample. In contrast, ETGs in non-isolated CGs show a higher fraction of quenched galaxies than isolated CGs and the control sample. Moreover, they exhibit lower sSFR in high stellar mass bins ($log(M_*/M_{\odot}) > 11$) relative to the control sample, underscoring the environmental effect in reducing star formation in this morphological type.

    \item The results summarised above motivated us to propose an evolutionary scenario for CGs, considering their connection to major structures. In this scenario, the major structures where CGs are embedded enhance the quenching process in CG galaxies and their morphological transformation (see Figure \ref{fig:esquema}). This, subsequently, influences the CG dynamics because non-isolated CGs have higher velocity dispersions than the isolated ones, suggesting that the dynamics of non-isolated CGs is dominated by the major structure where the CGs are located.

\end{enumerate}

These results add valuable evidence in support of CGs as places of galaxy transformation and evolution, providing the opportunity to develop future research of the environments of CGs. We highlight with this work the importance of carrying out detailed studies of the environment in which the CGs are found in order to further understand galaxy evolution in these systems.  As a follow-up, we plan to use 3D data from MaNGA (\citealt{2017wake}) to study the physical and kinematic properties of some of the galaxies in our sample, in order to understand in detail the physical phenomena that are occurring in this sample of galaxies in CGs.

\begin{acknowledgements}

We would like to thank the anonymous referee for their comments, as they helped improve the paper. G.P.M acknowledges financial support from ANID/'Beca de Doctorado Nacional'/21202024. G.P.M, A.M., F.A.G., and R.D gratefully acknowledge support by the ANID BASAL project FB210003, and funding from the Max Planck Society through a “PartnerGroup” grant. G.P.M and A.M acknowledge support by the FONDECYT Regular grant 1212046. F.A.G acknowledge support by the FONDECYT Regular grant 1211370. ST-F acknowledges the financial support of ULS/DIDULS through a regular project number PR2453858. MG acknowledges support from FAPERJ grant n. E-26/211.370/2021. DEO-R acknowledges the financial support from the Chilean National Agency for Research and Development (ANID), InES-Género project INGE210025. SP acknowledges the financial support of the Conselho Nacional de Desenvolvimento Científico e Tecnológico (CNPq) Fellowships 300936/2023-0 and 301628/2024-6.

The S-PLUS project, including the T80-South robotic telescope and the S-PLUS scientific survey, was founded as a partnership between the Fundação de Amparo à Pesquisa do Estado de São Paulo (FAPESP), the Observatório Nacional (ON), the Federal University of Sergipe (UFS), and the Federal University of Santa Catarina (UFSC), with important financial and practical contributions from other collaborating institutes in Brazil, Chile (Universidad de La Serena), and Spain (Centro de Estudios de Física del Cosmos de Aragón, CEFCA). We further acknowledge financial support from the São Paulo Research Foundation (FAPESP), the Brazilian National Research Council (CNPq), the Coordination for the Improvement of Higher Education Personnel (CAPES), the Carlos Chagas Filho Rio de Janeiro State Research Foundation (FAPERJ), and the Brazilian Innovation Agency (FINEP).

The authors are grateful for the contributions of CTIO staff in helping in the construction, commissioning and maintenance of the T80-South telescope and camera. We are also indebted to Rene Laporte and INPE, as well as Keith Taylor, for their important contributions to the project. We also thank CEFCA staff for their help with T80-South. Specifically, we thank Antonio Marín-Franch for his invaluable contributions in the early phases of the project, David Cristóbal-Hornillos and his team for their help with the installation of the data reduction package jype version 0.9.9, César Íñiguez for providing 2D measurements of the filter transmissions, and all other staff members for their support.
\end{acknowledgements}


\bibliographystyle{aa} % style aa.bst
\bibliography{example} % your references Yourfile.bib

\end{document}